\documentclass[journal,comsoc]{IEEEtran}
\usepackage[T1]{fontenc}

\usepackage[spaces,obeyspaces]{url}
\makeatletter
\g@addto@macro{\UrlBreaks}{\UrlOrds}
\makeatother

\usepackage{pgfplots}
\usepgfplotslibrary{polar}
\usepackage{gensymb}
\usepackage{booktabs}
\definecolor{red}{rgb}{0.8,0.145,0.161}
\definecolor{blue}{rgb}{0.224,0.416,0.694}
\definecolor{green}{rgb}{0.243,0.588,0.318}
\definecolor{grey}{rgb}{0.326,0.318,0.329}
\definecolor{orange}{rgb}{0.855,0.486,0.188}

\ifCLASSINFOpdf
\usepackage{amsmath}
\ifCLASSOPTIONcompsoc
  \usepackage[caption=false,font=normalsize,labelfont=sf,textfont=sf, justification=centering]{subfig}
\else
  \usepackage[caption=false,font=footnotesize, justification=centering]{subfig}
\fi

\usepackage{import}
\usepackage{multirow}
\usepackage{array}
\usepackage{cite}

\usepackage{pifont}
\newcommand{\cmark}{\color{green} \ding{51}}%
\newcommand{\xmark}{\color{red} \ding{55}}%
 
\usepackage[cmintegrals]{newtxmath}

\begin{document}

\title{Urban Outdoor Measurement Study of Phased Antenna Array Impact on Millimeter-Wave Link Opportunities and Beam Misalignment}

\author{Lars~Kuger,~\IEEEmembership{}%
        Aleksandar~Ichkov,~\IEEEmembership{}%
        Petri~M\"ah\"onen~\IEEEmembership{}%
        and~Ljiljana~Simi\'c~\IEEEmembership{}
\thanks{All authors are with the Institute for Networked Systems, 
RWTH Aachen University, Kackertstrasse 9, 52072 Aachen, Germany.}
\thanks{E-mail: \{lku, aic, pma, lsi\}@inets.rwth-aachen.de.}
\thanks{}
\thanks{}}



\maketitle


\begin{abstract}
Exploiting multi-antenna technologies for robust beamsteering to overcome the effects of blockage and beam misalignment is the key to providing seamless multi-Gbps connectivity in millimeter-wave (mm-wave) networks. In this paper, we present the first large-scale outdoor mm-wave measurement study using a phased antenna array in a typical European town. We systematically collect fine-grained 3D angle-of-arrival (AoA) and angle-of-departure (AoD) data, totaling over 50,000 received signal strength measurements. We study the impact of phased antenna arrays in terms of number of link opportunities, achievable data rate and robustness under small-scale mobility, and compare this against reference horn antenna measurements. Our results show a limited number of 2--4 link opportunities per receiver location, indicating that the mm-wave multipath richness in a European town is surprisingly similar to that of dense urban metropolises. The results for the phased antenna array reveal that significant losses in estimated data rate occur for beam misalignments in the order of the half-power beamwidth, with significant and irregular variations for larger misalignments. By contrast, the loss for horn antennas is monotonically increasing with the misalignment. Our results strongly suggest that the effect of non-ideal phased antenna arrays must be explicitly considered in the design of agile beamsteering algorithms.

\end{abstract}

%
\IEEEpeerreviewmaketitle

\section{Introduction}

Multi-antenna technologies are the key enabler for unlocking the potential of millimeter-wave (mm-wave) spectrum bands in 5G-and-beyond networks~\cite{5G, xiao_millimeter_2017}. A number of mm-wave outdoor measurement campaigns, predominantly using channel sounder setups with horn antennas, e.g.~\cite{rappaport_wideband_2015,samimi_28_2013, simic_60_2016, weiler_measuring_2014, rajagopal_channel_2012, ko_millimeter-wave_2017}, have now demonstrated the fundamental feasibility of mm-wave links in urban environments and resulted in mm-wave statistical channel models (see \cite{rappaport_overview_2017} and references therein). This year has also seen the first test commercial mm-wave outdoor deployments, albeit with limited capabilities and performance~\cite{Verizon_5G}. Directional high-gain antenna beams are used to overcome the high path loss at mm-wave frequencies, but the sensitivity to misalignment of antenna beams \cite{simic_60_2016,lee_field-measurement-based_2018} and the effect of link blockage due to environmental and mobile obstacles such as buildings \cite{PenetrationLoss_2013} and humans \cite{Qualcomm_2019}, remain primary challenges for seamless coverage in large-scale mobile network deployments. The key to making mm-wave cellular networks a reality is thus exploiting multi-antenna technology for robust and precise beamsteering to overcome effects of blockage and beam misalignment caused by large and small-scale mobility, and provide seamless gigabit-per-second (Gbps) connectivity.

Great advances have been reported in the literature regarding mm-wave antenna technology and testbeds \cite{aryanfar_millimeter-wave_2015, sadhu_128-element_2018, aziz_high_2018, saha_x60_2017}, but the evaluation of these devices has been largely conducted in controlled indoor environments.
The notable exceptions to this are~\cite{Qualcomm_2018, roh_millimeter-wave_2014,kurita_outdoor_2019, MiWeBa}, which demonstrated outdoor coverage measurements using mm-wave phased antenna arrays. However, these works lack fine-grained angle-of-arrival (AoA) and angle-of-departure (AoD) measurements, which are crucial for understanding the impact of beam misalignment on mm-wave link and network performance, and for evaluation of beam training algorithms that are essential for mm-wave initial access and mobility management. Other prior outdoor measurement studies, e.g.~\cite{rappaport_wideband_2015,samimi_28_2013, simic_60_2016, weiler_measuring_2014, rajagopal_channel_2012, ko_millimeter-wave_2017}, used horn antenna or omni-directional antenna setups. The existing literature thus offers very limited insight on the performance of mm-wave antenna arrays in a real outdoor urban network environment, in particular with respect to the beamsteering opportunities and beam misalignment effects, which is a crucial input for system-level engineering design of future mm-wave networks.

In this paper, we present the first large-scale outdoor urban mm-wave measurement study using a state-of-the-art phased antenna array, collecting received signal strength (RSS) data over systematic fine-grained 3D AoA and AoD orientations in a typical European town. We study the impact of phased antenna arrays in terms of the number of link opportunities, achievable data rate and robustness under small-scale mobility -- which we emulate based on our fine-grained angular measurements -- and directly compare this against reference horn antenna measurements. Our results show a limited number of 2--4 available distinct spatial link opportunities per receiver location, indicating that the mm-wave multipath richness in a typical European town center is surprisingly similar to that in dense urban areas as presented in~\cite{rappaport_wideband_2015,samimi_28_2013, weiler_measuring_2014, ko_millimeter-wave_2017}. The results for the phased antenna array reveal that losses in estimated data rate of up to 70\% occur for small beam misalignments in the order of the half-power beamwidth (HPBW), with significant and irregular variations in estimated data rate for larger misalignments due to the non-ideal phased antenna array radiation pattern. By contrast, the loss in estimated data rate for the horn antenna setup is monotonically increasing with the orientation error. This shows that earlier measurement studies \cite{simic_60_2016,lee_field-measurement-based_2018} or theoretical studies \cite{wildman_joint_2014} of beam misalignment effects on mm-wave link and network performance using horn antennas or idealized directional antenna patterns cannot be simply generalized. Consequently, we analyze the implications that our results have on the design for future beamsteering algorithms in mm-wave outdoor network deployments.

The rest of the paper is organized as follows. Sec.~\ref{sec:related_work} gives the related work overview. Sec.~\ref{sec:setup} presents our measurement setup and methodology. Our measurement results are presented and analyzed in Sec.~\ref{sec:results}. In Sec.~\ref{sec:discussion} we discuss the engineering implication of our findings for outdoor mm-wave network deployments. Finally, Sec.~\ref{sec:conclusion} concludes the paper.

\section{Related Work} \label{sec:related_work}

Recent years have seen great advances in mm-wave phased antenna arrays and corresponding transceivers for base stations (BSs) and user equipment (UE). For instance, Samsung \cite{aryanfar_millimeter-wave_2015}, IBM and Ericsson \cite{sadhu_128-element_2018}, and Sivers IMA \cite{aziz_high_2018} have presented mm-wave phased antenna array modules, and new mm-wave testbed using e.g. SiBeam's phased antenna arrays have been demonstrated \cite{saha_x60_2017}. However, these new mm-wave transceivers and phased antenna array designs have largely been evaluated in controlled environments \cite{aryanfar_millimeter-wave_2015, sadhu_128-element_2018, saha_x60_2017, aziz_high_2018}, rather than in real outdoor urban settings. Notable exceptions to this are the studies in \cite{Qualcomm_2018, roh_millimeter-wave_2014,kurita_outdoor_2019, MiWeBa}. In \cite{Qualcomm_2018, roh_millimeter-wave_2014, kurita_outdoor_2019} data rate measurements with 28 GHz multi-antenna arrays were conducted in outdoor measurements, analyzing the possible coverage in potential mm-wave cells. In \cite{MiWeBa} the RSS over AoD was measured at several receiver positions in a car park using an omni-directional antenna at the receiver and a 60 GHz phased antenna array at the transmitter, finding strong reflected paths for many receiver positions. Yet, these studies did not include detailed AoA \emph{and} AoD measurements to thoroughly investigate beamsteering opportunities and limitations caused by beam misalignment. Table~\ref{tab:relwork} shows an overview of related mm-wave measurements. By contrast, in this work we present the results of the first comprehensive large-scale urban measurement campaign with mm-wave phased antenna arrays systematically collecting RSS over fine-grained AoA and AoD antenna orientations.

{
	\begin{table*}[t]
		\centering
		\caption{Overview of related work showing the used antenna type, whether AoA and AoD was collected  systematically outdoors (out.) with angular azimuth resolution $\Delta$, the maximum measured TX-RX beam orientation pairs per TX-RX location $N$, and whether beam misalignment (mis.) was studied.\label{tab:relwork}} 
		\begin{tabular}{l c c c c c c c c c c} \toprule
			\multirow{2}{*}{Ref.} 											& \multirow{2}{*}{Out.} & \multirow{2}{*}{Environment}  & \multicolumn{3}{c}{Antenna type} &  \multirow{2}{*}{AoA} & \multirow{2}{*}{AoD} & \multirow{2}{*}{$\Delta$} &  \multirow{2}{*}{$N$} &\multirow{2}{*}{Mis.} \\ 
			& & & Omni & Horn & Array & & & & \\	\midrule 
			\cite{sadhu_128-element_2018, aziz_high_2018, aryanfar_millimeter-wave_2015} & \xmark & Lab & -- & \xmark & \cmark  &  \xmark & \xmark & N/A & N/A & \xmark \\
			\cite{weiler_measuring_2014} 					& \cmark & Large European city & \cmark & \xmark & \xmark &  \xmark & \xmark & N/A & N/A & \xmark \\
			\cite{Qualcomm_2018, roh_millimeter-wave_2014, kurita_outdoor_2019} 		& \cmark & Parking lot/campus/Tokyo & -- & \xmark & \cmark  &  \xmark & \xmark & N/A & N/A & \xmark \\
			\cite{MiWeBa} 									& \cmark & Parking lot & \cmark & \xmark & \cmark &  \xmark & \cmark & $5^\circ$ & 133 & \xmark \\
			\cite{rajagopal_channel_2012} 					& \cmark & Parking lot & -- & \cmark & \xmark &  \cmark & \xmark & $10^\circ$  & 108 & \xmark \\
			\cite{lee_field-measurement-based_2018}			 & \cmark & Large Korean city & \cmark & \cmark & \xmark &  \cmark & \xmark & $10^\circ$ & 36 & \cmark\\
			\cite{ko_millimeter-wave_2017} 					& \cmark & Large Korean city & -- & \cmark & \xmark &  \cmark & \cmark & $10^\circ$  & 36,504 & \xmark\\
			\cite{rappaport_wideband_2015, samimi_28_2013} & \cmark  & US cities & -- & \cmark & \xmark &  \cmark & \cmark & 5--10$^\circ$ & 540 & \xmark\\
			\cite{saha_x60_2017} 							& \xmark & Office & -- & \xmark & \cmark &  \cmark & \cmark & $5^\circ$ & 625 & \cmark \\
			\cite{simic_60_2016}			 				& \cmark & Small European city & -- & \cmark & \xmark &  \cmark & \cmark & $3.6^\circ$  & 64,000 & \cmark \\
			\textbf{This paper}										& \cmark & \textbf{Small European city} & -- & \cmark & \cmark &  \cmark & \cmark & \textbf{5--9$^\circ$} & \textbf{7,200} & \cmark  \\ \bottomrule 
		\end{tabular}
	\end{table*}
}

Independently from the advances in mm-wave multi-antenna technology, a number of mm-wave measurement campaigns with horn or omnidirectional antennas have been conducted in outdoor environments with the aim of establishing \emph{statistical} mm-wave channel models (see \cite{xiao_millimeter_2017} and references therein). Rappaport \emph{et al.} \cite{rappaport_wideband_2015} conducted seminal measurements in urban areas using a channel sounder with mechanically steerable horn antennas to record power delay profiles for a limited selected range of AoA and AoD combinations at each transmitter~(TX)/receiver~(RX) pair. Similar measurement methodologies have been employed in other measurement campaigns, e.g. \cite{samimi_28_2013, weiler_measuring_2014, rajagopal_channel_2012, ko_millimeter-wave_2017}, and measurement-based statistical channel models such as the NYU model \cite{samimi_3-d_2016} have consequently been proposed (see \cite{rappaport_overview_2017} and references therein). Overall, the studies in \cite{rappaport_wideband_2015, samimi_28_2013, weiler_measuring_2014, ko_millimeter-wave_2017} reported a limited number of 2--5 mm-wave multipath clusters per TX/RX pair for urban areas in different metropolises. However, as these measurement campaigns used horn or omnidirectional antennas and focused on statistical channel characteristics -- rather than on systematic RSS over AoA/AoD measurements as in our work -- they do not allow us to study the impact of phased antenna arrays on spatial link opportunities and beam misalignment in mm-wave networks. By contrast, we conducted our measurement campaign with both phased antenna arrays and horn antennas to bridge the gap between prior measurement campaigns using horn antennas \cite{rappaport_wideband_2015, samimi_28_2013, rajagopal_channel_2012, ko_millimeter-wave_2017} and future mm-wave networks using phased antenna arrays. Moreover, we conducted measurements in a typical European town with significantly different building layout and materials as compared to big cities as in most prior outdoor studies (\emph{cf.} Table~\ref{tab:relwork}).

Lee \emph{et al.} \cite{lee_field-measurement-based_2018} studied the effect of beam misalignment based on measurements taken with horn antennas. To that end, mm-wave power measurements were systematically collected across azimuth angles in both free-space and urban environments. The conclusion based on these horn antenna measurements was that for a fixed beamwidth, the power loss increases linearly with increasing beam misalignment until saturation is reached. This confirms the results of Simi\'c \emph{et al.} in \cite{simic_60_2016}, where the RSS over AoA/AoD was systematically measured with a horn antenna setup in an urban environment in the 60 GHz band. The results in \cite{simic_60_2016,lee_field-measurement-based_2018} stand in contrast to the results of our phased antenna array measurements in this paper, which show a distinct non-linear relationship between misalignment and power loss. Moreover, our results thereby clarify artifacts that were observed in \cite{saha_x60_2017}, where a mm-wave testbed using a phased antenna array with 25$^\circ$ beams was used to study the effect of beam misalignment on the SNR in indoor environments. For large misalignments the results showed artifacts, which the authors speculated would not occur with horn antennas. In our work, we demonstrate that this indeed is caused by the non-ideal, phased antenna array pattern by explicitly comparing our measurements against those with a horn antenna in the same outdoor scenarios. Our comparison underlines the paradigm shift that occurs in terms of beam misalignment effects -- and thus practical beam training implications -- when moving from horn antennas to real phased antenna arrays.

We emphasize that while a large number of beam training strategies for initial random access and mobility management in mobile mm-wave networks have been proposed, their evaluation has likewise been largely restricted to measurements with horn antenna setups \cite{sur_beamspy:_2016}, simulations based on simplified beam models \cite{giordani_comparative_2016}, or commercial off the shelf equipment with limited control of the actual beam patterns~\cite{steinmetzer_compressive_2017}. Yet, for practical testing and development of high-performing beam training algorithms, rich angular data based on real phased antenna array measurements as presented in this paper is essential. To this end, we provide open access to our data~\cite{inets_data}.


In addition to empirical research, the effect of beam misalignment on mm-wave network performance has also been addressed from a theoretical perspective. For instance, Wildman \emph{et al.} \cite{wildman_joint_2014} show that in theory sidelobes can be beneficial to the success probability of a transmission in low density networks and that the spatial throughput and transmission capacity maximizing beamwidth has a nearly linear relationship with the mean orientation error for Gaussian and sectored antenna radiation pattern models. As such idealized radiation pattern models are much more similar to a horn antenna radiation pattern than that of a real phased antenna array, our results strongly suggest that a linear relationship will \emph{not} hold in real mm-wave deployments and that the consequences of this idealization on the requirements and opportunities of beamsteering algorithms and network system design have been underestimated. Therefore, our work forms a basis for realistic models of beam misalignment effects in urban mm-wave networks.

\section{Measurement Setup \& Methodology} \label{sec:setup}

Our outdoor measurement campaign was conducted in the German town of Langenfeld during the summer of 2018. The study area constitutes the central pedestrian zone with mainly shops in the surroundings. The buildings typically have 2-5 floors and are constructed of concrete with large windows and some metal parts on the facades. 

\begin{figure}
	\centering
	\subfloat[TX 1 seen from RX A.\label{subfig:txrxposphotos_tx1rxa}]{
		\includegraphics[width=2.6cm, height=2.6cm]{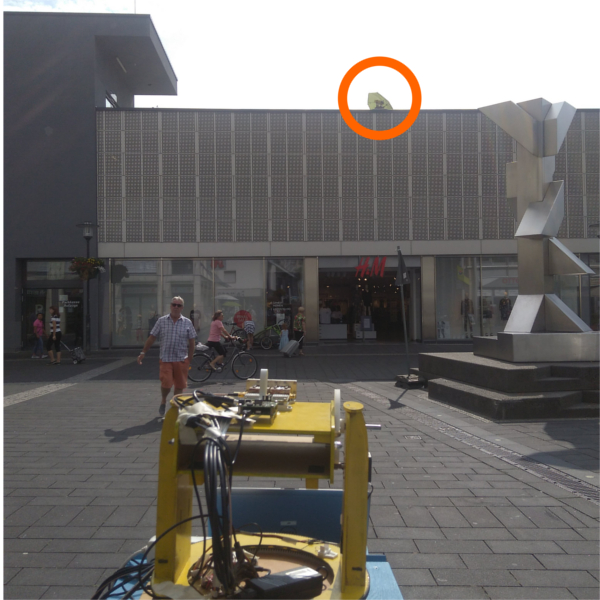}
	} 
	\subfloat[TX 1 and RX B.\label{subfig:txrxposphotos_tx1rxb}]{
		\includegraphics[width=2.6cm, height=2.6cm]{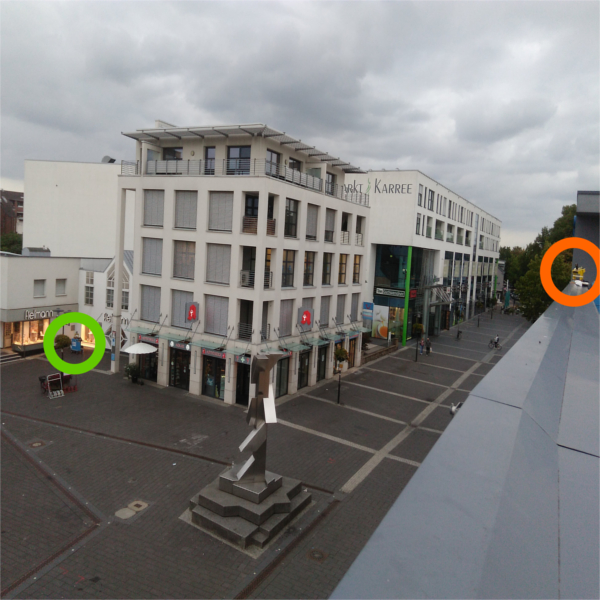}
	} 
	\subfloat[RX B seen from TX 2.\label{subfig:txrxposphotos_tx2rxb_tx}]{
		\includegraphics[width=2.6cm, height=2.6cm]{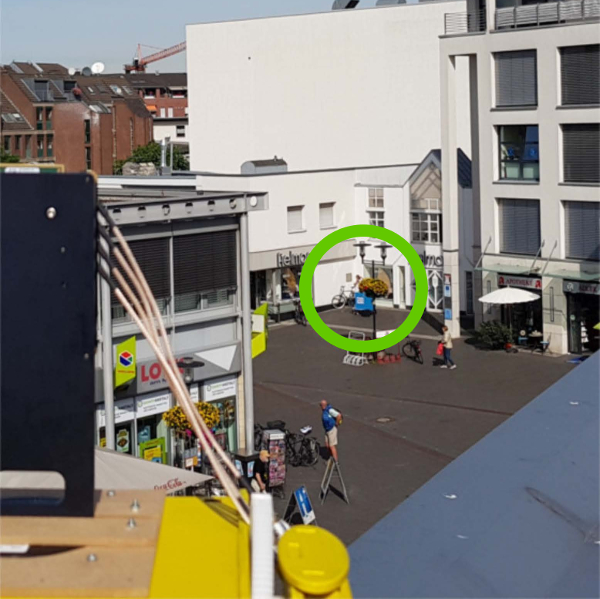}
	}\\
	\subfloat[RX C seen from TX 1.\label{subfig:txrxposphotos_tx1rxc_tx}]{
		\includegraphics[width=2.6cm, height=2.6cm]{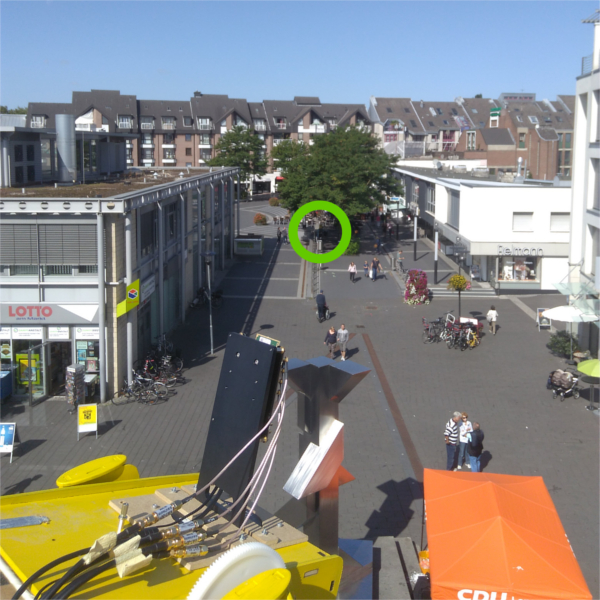}
	} 
	\subfloat[TX 2 and RX D.\label{subfig:txrxposphotos_tx2rxd}]{
		\includegraphics[width=2.6cm, height=2.6cm]{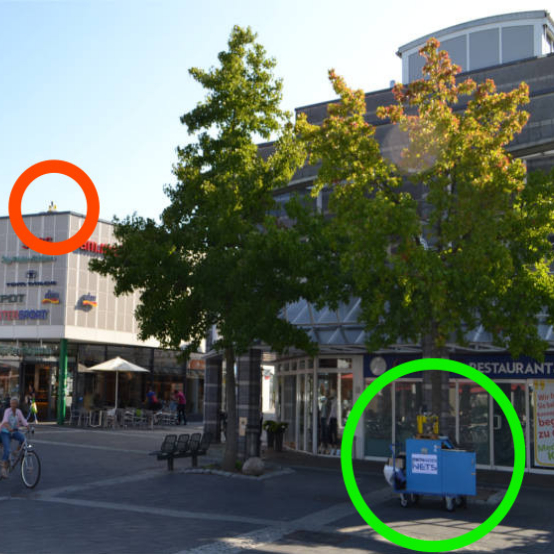}
	}
	\subfloat[3D turntable with mounted antenna array transceiver\label{subfig:turntable}]{
		\includegraphics[width=2.6cm, height=2.6cm]{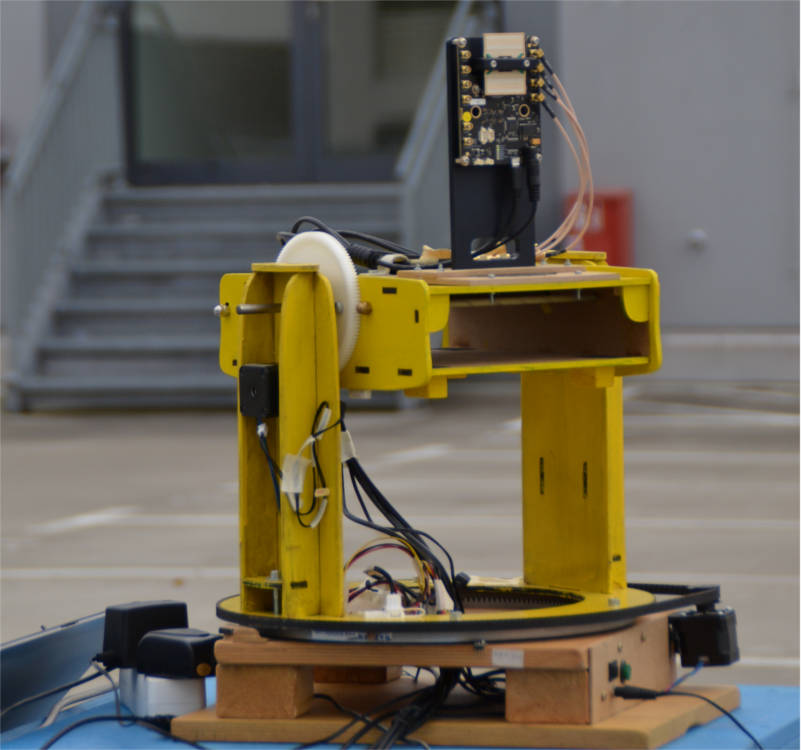}
	} 
	\caption{TX and RX in different positions during measurements. The RX is circled in green and TX in orange.\label{fig:txrxposphotos}}
\end{figure}

The TX was located at different positions on the rooftop of a multi-floor car park at a height of 9.7~m. The RX positions A--D were located in the pedestrian zone, i.e. on ground level, with the mounted antenna at a height of 1.7~m. Fig.~\ref{fig:langenfeldtxrxpos} shows an aerial image of the area where the TX positions (1, 2) and RX positions (A, B, C, D) are marked. Fig.~\ref{fig:txrxposphotos} shows the different TX/RX positions during the measurements. The RX is circled in green and the TX in orange. The positions were chosen based on representative scenarios as follows:

\begin{figure}
	\centering
	\includegraphics[scale=0.23]{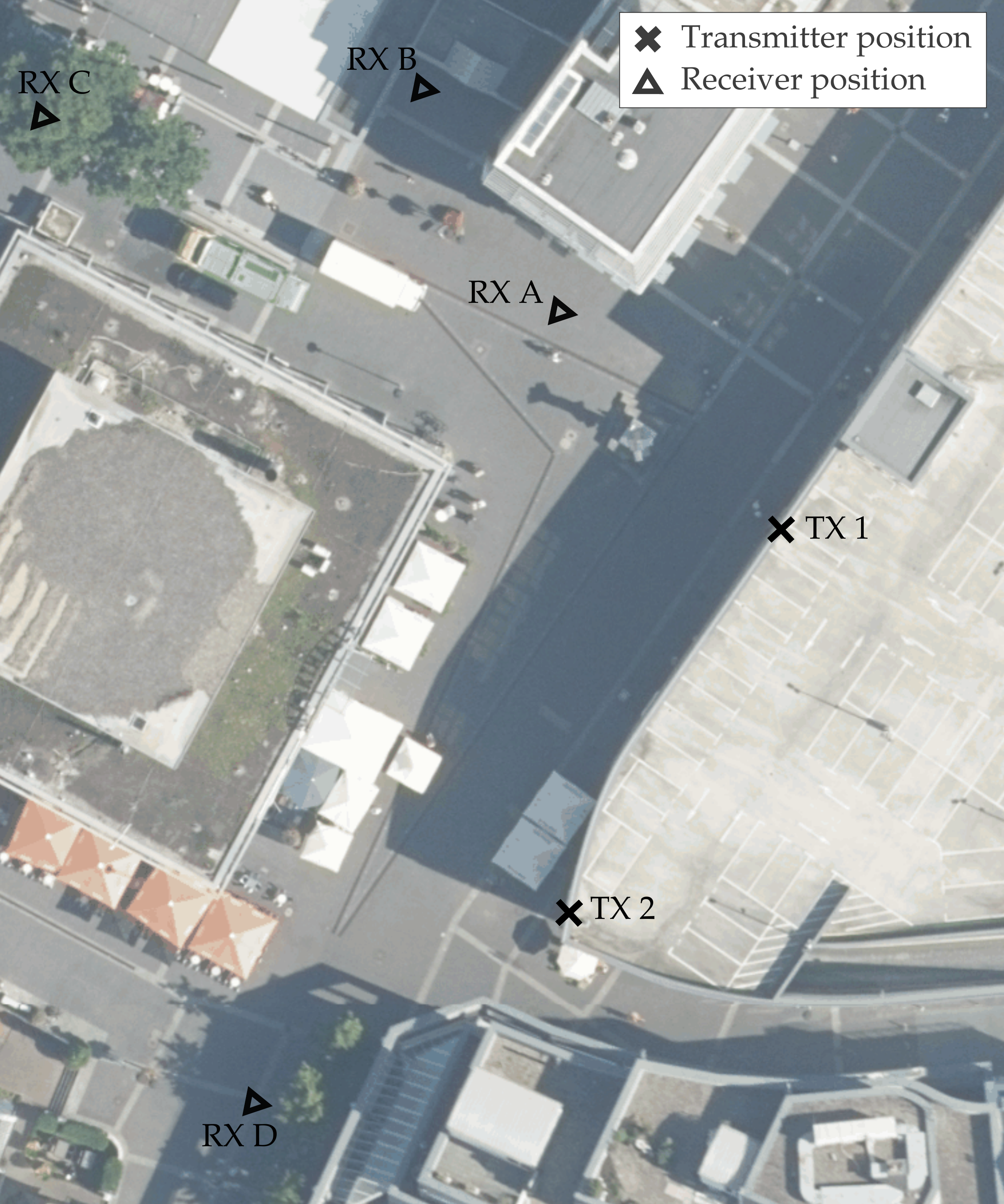}
	\caption{Aerial view of the TX (1-2) and RX positions (A-D). The measurement scenarios are listed in Table~\ref{tab:measurements}. (Photo courtesy of \cite{land_nrw_tim-online_2019}).}
	\label{fig:langenfeldtxrxpos}
\end{figure}

\begin{itemize}
	\item 
	\textit{TX1-RXA}: Typical scenario with clear LOS and TX-RX distance of 25~m (Fig.~\ref{subfig:txrxposphotos_tx1rxa}).
	
	\item 
	\textit{TX1-RXB \text{and} TX2-RXB}: RX-B was chosen to investigate the effects of an obstructed LOS and evaluate the coverage from different TX positions for the same RX position. In scenario TX1-RXB the LOS is partially blocked by a building corner, while in scenario TX2-RXB the LOS is partially blocked by a lamp-post with a flower box mounted on it (Figs.~\ref{subfig:txrxposphotos_tx1rxb}-\ref{subfig:txrxposphotos_tx2rxb_tx}). 
	
	\item
	\textit{TX1-RXC}: Represents a typical street canyon with 18~m width and buildings to either side. The LOS path was predominantly clear despite tree foliage \mbox{(Fig.~\ref{subfig:txrxposphotos_tx1rxc_tx})}.
	
	\item 
	\textit{TX2-RXD}: Scenario chosen to investigate the effect of tree foliage on mm-wave coverage, with a TX-RX distance of 28~m and clearly visible trees obstructing the LOS \mbox{(Fig.~\ref{subfig:txrxposphotos_tx2rxd})}.
\end{itemize}

For our measurement campaign we used two different 60~GHz transceiver setups, one using a phased antenna array and the other using a horn antenna for comparison. A relatively narrowband signal transmission was chosen (1~MHz) to obtain fine-grained RSS over AoA/AoD data, allowing us to have a robust setup
that does not require extensive calibration as the gain over the
frequency band is flat. Since we are not interested in a time-characterization of the channel, the narrowband power measurements are sufficient for our objective of characterizing spatial mm-wave link opportunities in a typical outdoor environment and beamsteering requirements for maintaining a reliable link connectivity under small-scale mobility. We note that frequency bands in the range of 24--86~GHz are under consideration for 5G-and-beyond mm-wave cellular networks~\cite{ITU_WRC-15}. In this paper, we report measurements in the 60~GHz band without the loss of generality and for comparability with prior mm-wave outdoor urban measurement studies, e.g.~\cite{rappaport_wideband_2015, simic_60_2016, Qualcomm_2018,MiWeBa}.

\subsection{Phased Antenna Array Setup}
\label{sec:PhasedAntennaSetup}

\begin{figure}[tb]
	\centering
	\hspace{3ex}
	\subfloat[Phased antenna array board with $a=0.71$mm, $y=2.4$mm, $x=2.97$mm. \label{subfig:16x2array}]{
		\centering
		\includegraphics[width=0.16\textwidth]{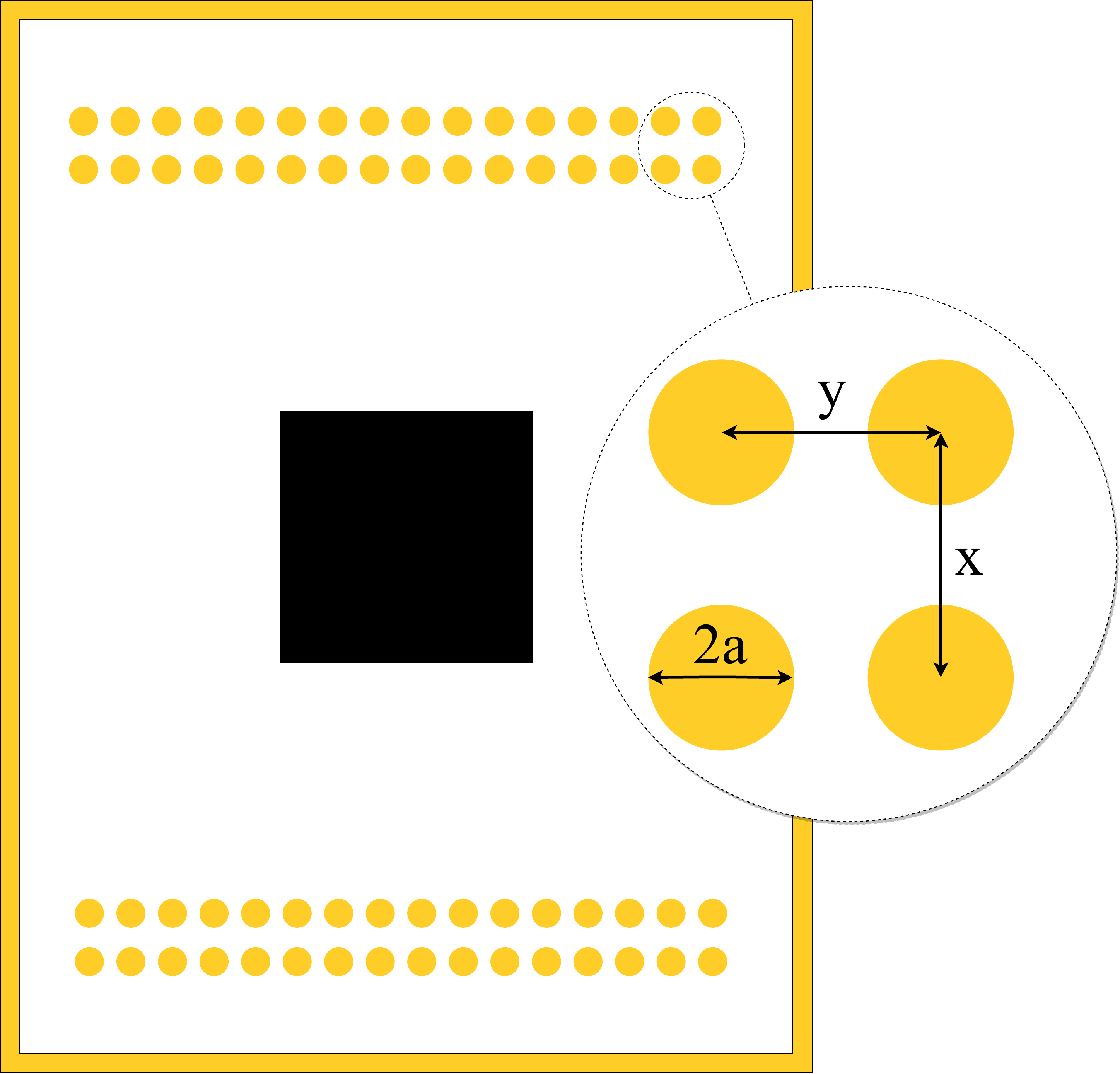}	
	}
	\hspace{3ex}
	\subfloat[The simulated horn antenna power pattern.\label{subfig:FM25240_HE} ]{
		\begin{tikzpicture}[scale=0.35]
\begin{polaraxis}[
   xticklabel=$\pgfmathprintnumber{\tick}^\circ$,
   xtick={0, 30, ..., 330}, 
   ytick={-40, -30, ...,20},
   ymin=-40, ymax=20,
   y coord trafo/.code=\pgfmathparse{#1+40},
   rotate=-90,
   y coord inv trafo/.code=\pgfmathparse{#1-40},
   x dir=reverse,
   xticklabel style={anchor=-\tick-90, font=\Large},
   yticklabel style={anchor=east, xshift=6.6cm, font=\Large},
   y axis line style={yshift=4.5cm},
   ytick style={yshift=4.5cm},
   yticklabel=$\pgfmathprintnumber{\tick}\,\mbox{dBi}$,
   legend style={at=(current axis.south east), anchor=south west, yshift=-2.5cm, xshift=2.2cm, font=\Large}
]

\addplot [no markers, thick, solid, red] table [col sep=comma] {data/FM25240_Hor.csv};
\addlegendentry{H-plane};
\addplot [no markers, thick, orange, dashed, line width=0.8mm] table [col sep=comma] {data/FM25240_El.csv};
\addlegendentry{E-plane};

\end{polaraxis}
\end{tikzpicture}
	} \\
	\subfloat[The phased antenna array H-plane power pattern normalized\newline to unity at the maximum.\label{subfig:16x2patternH}]{
		\begin{tikzpicture}[scale=0.35]
\begin{polaraxis}[
   xticklabel=$\pgfmathprintnumber{\tick}^\circ$,
   xtick={0, 30, ..., 330}, 
   ytick={-50, -40,...,0},
   ymin=-50, ymax=0,
   y coord trafo/.code=\pgfmathparse{#1+50},
   rotate=-90,
   y coord inv trafo/.code=\pgfmathparse{#1-50},
   x dir=reverse,
   xticklabel style={anchor=-\tick-90, font=\Large},
   yticklabel style={anchor=east, xshift=6.6cm, font=\Large},
   y axis line style={yshift=4.5cm},
   ytick style={yshift=4.5cm},
   yticklabel=$\pgfmathprintnumber{\tick}\,\mbox{dB}$,
   legend style={at=(current axis.south east), anchor=south west, yshift=-2.5cm, xshift=1.4cm, font=\Large}
]
\addplot [no markers, thick, dashed, blue, line width=0.8mm] table [col sep=comma] {data/20180702-Run3-20180703-Run4.csv};
\addlegendentry{Measurement};
\addplot [no markers, thick, red] table [col sep=comma] {data/AntPatHor_58G.csv};
\addlegendentry{Simulation};

\end{polaraxis}
\end{tikzpicture}
	}  
	\subfloat[The phased antenna array E-plane power pattern normalized to unity at the maximum.\label{subfig:16x2patternEl}]{
		\begin{tikzpicture}[scale=0.35]
\begin{polaraxis}[
   xticklabel=$\pgfmathprintnumber{\tick}^\circ$,
   xtick={0,30,...,330},
   ytick={-50, -40,...,0},
   ymin=-50, ymax=0,
   y coord trafo/.code=\pgfmathparse{#1+50},
   rotate=-90,
   y coord inv trafo/.code=\pgfmathparse{#1-50},
   x dir=reverse,
   xticklabel style={anchor=-\tick-90, font=\Large},
   yticklabel style={anchor=east, xshift=6.6cm, font=\Large},
   y axis line style={yshift=4.5cm},
   ytick style={yshift=4.5cm},
   yticklabel=$\pgfmathprintnumber{\tick}\,\mbox{dB}$,
   legend style={at=(current axis.south east), anchor=south west, yshift=-2.5cm, xshift=1.4cm, font=\Large}
]
\addplot [no markers, thick, dashed, blue, line width=0.8mm] table [col sep=comma] {data/20180731_El_Pattern.csv};
\addlegendentry{Measurement};
\addplot [no markers, thick, red] table [col sep=comma] {data/AntPatEl_58G.csv};
\addlegendentry{Simulation};
\end{polaraxis}

\end{tikzpicture}
	} 
	\caption{Schematic drawing of the Sivers IMA TRX BF/01 phased antenna array board and antenna radiation patterns.}
\end{figure}
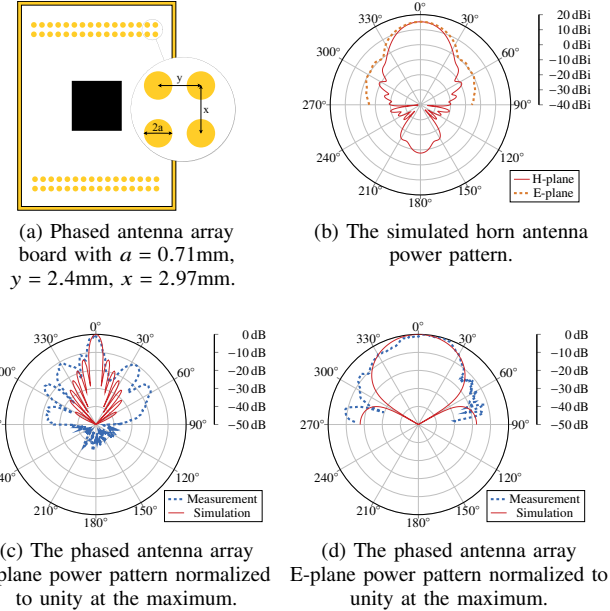

We used the SiversIMA 57-71~GHz phased antenna array radio frequency integrated circuit (RFIC) TRX BF/01 \cite{siversima_RFIC} with the corresponding evaluation kit EVK06002 \cite{SiversIMATRX}. The evaluation kit includes a 16+16 (TX/RX) patch antenna module \cite{aziz_high_2018} and based on a codebook with 64 different entries, the beam can be steered in the azimuth plane in the range $[+45^\circ, -45^\circ]$. Fig.~\ref{subfig:16x2array} shows a schematic drawing of the phased antenna array board and the RFIC. Figs.~\ref{subfig:16x2patternH}-\ref{subfig:16x2patternEl} compare the measured phased antenna array radiation pattern with the simulated one obtained using \textsc{Matlab}'s Phased Array System Toolbox based on generic patch antenna elements. We note the considerable differences between the ideal, simulated and the real, measured radiation pattern.

\begin{figure*}[t]
	\centering
	\includegraphics[width=0.78\textwidth]{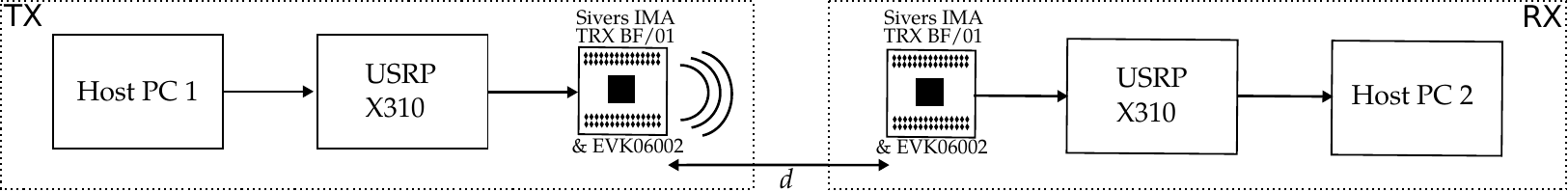}
	\caption{Schematic of the phased array setup. Distance $d$ depends on the TX/RX locations.\label{fig:antarraysetup}}
	
	 \vspace*{\floatsep}
	\centering
	\includegraphics[width=1\textwidth]{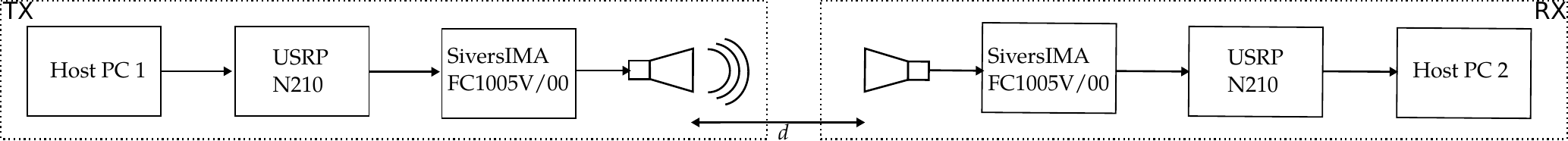}
	\caption{Schematic of the horn antenna setup. Link distance $d$ depends on the TX/RX locations.}
	\label{fig:hornantennasetup}
\end{figure*}

Fig.~\ref{fig:antarraysetup} shows the schematic of the phased antenna array setup with the TX including host PC 1, USRP X310 SDR with LFTX daughterboards, and the phased antenna array transceiver, and the RX consisting of an equivalent setup. At the TX, the host PC 1 generates a complex sinusoidal signal at the frequency $f_{sin}=250$~kHz, which the USRP converts to an analog baseband signal with a sampling rate $f_s=1$~MHz. The phased antenna array transceiver upconverts the signal to a frequency $f_c=58.32$~GHz, and transmits the signal over the air. The RX transceiver downconverts the signal before it is sampled with sampling rate $f_s$ at the USRP. At the host PC 2, the power spectrum of the signal is computed using a flattop window. Finally, the signal strength $RSS_{A,meas}$ of the received sinusoidal signal is extracted from the power spectrum. This was calibrated against a setup of standard gain horn antenna, Agilent 11970V mixer, and Agilent E4440A spectrum analyzer with a resolution bandwidth similar to the binwidth of the digitally computed power spectrum; differences in antenna gain and cable losses were accounted for.

\subsection{Horn Antenna Setup}
\label{sec:HornAntennaSetup}

For comparison, our second setup used the FC1005V/00 57-66~GHz converter by SiversIMA \cite{SiversIMAVband} and a standard 15~dBi gain horn antenna \cite{Flann}. Fig.~\ref{subfig:FM25240_HE} shows the corresponding simulated radiation pattern. The schematic of the horn antenna setup is shown in Fig.~\ref{fig:hornantennasetup}. The left part of the figure shows the TX including the host PC 1, USRP N210 with WBX daughterboard, Sivers IMA FC1005V/00 upconverter and the horn antenna. The right part of the figure shows the RX consisting of a similar setup as the TX. The transmission and reception signal chain of the horn antenna setup is equivalent to the phased antenna array setup (\emph{cf.}~Sec.~\ref{sec:PhasedAntennaSetup}), except that the signal between the USRP and up-/downconverter is at the intermediate frequeny $f_{IF}=1.32$~GHz and not at baseband. The RSS at the horn antenna setup $RSS_{H,meas}$ was obtained by the same calibrated procedure as described in Sec.~\ref{sec:PhasedAntennaSetup}.

\subsection{Measurement Procedure}

To identify potential link opportunities and their respective AoA and AoD, we systematically swept the angular space by changing the beam orientations at the TX and RX. 

\begin{table*}
	\centering
	\caption{Measurement scenarios with ID consisting of a number (TX position), capital letter (RX position) and small letter (setup: \emph{(a)} phased antenna array and \emph{(h)} horn antenna). The distance $d$ is the 3D link distance. $N_{meas, tot}$ is the total number of measured TX/RX beam orientation combinations.}
	\label{tab:measurements}
	{
		\begin{tabular}{c c  c c c c c c c c c}
			\toprule
			ID & TX & RX & $d$ (m) & Type & Setup & $\theta_{TX}(^\circ)$ & $\phi_{TX}(^\circ)$  & $\theta_{RX}(^\circ)$ & $\phi_{RX}(^\circ)$ & $N_{meas, tot}$ \\ 
			\midrule
			\textit{1Aa}  & 1	   & A      &  25.3       & LOS  & Array  &     -15     & -90:6:84  & -30:30:60  & -180:6:174  &  7200 \\
			\textit{1Ah}  & 1	   & A      &  25.3       & LOS  & Horn   &     -15     & -54:6:42  & 0:10:60  & -180:6:174  & 7140  \\
			\textit{1Ba}  & 1	   & B      &  43.8       & NLOS  & Array &     -15     & -90:6:84  & -30:30:60  & -180:6:174 & 7200   \\
			\textit{1Bh}  & 1	   & B      &  43.8       & NLOS  & Horn  &     -15     & -54:6:6   & 0:10:60  & -180:6:174  & 4620  \\
			\textit{1Ca}  & 1	   & C      &  64.9       & LOS  & Array  &     -15     & -90:6:84  & -30:30:60  & -180:6:174  & 7200  \\
			\textit{1Ch}  & 1	   & C      &  64.9       & LOS  & Horn   &     -15     & -24:6:30  & -30:10:60  & -180:6:174  & 6000  \\
			\textit{2Da}  & 2	   & D      &  27.8       & NLOS  & Array &     -15     & -72:6:52  & -30:30:60  & -180:6:174  & 5200  \\
			\textit{2Ba}  & 2	   & B      &  63.2       & NLOS  & Array&     -15     & -90:6:84  & -30:30:60  & -180:6:174  & 7200  \\
			\bottomrule
		\end{tabular}
	}
\end{table*}

\subsubsection{Phased Antenna Array Transceiver}

For the measurements, we covered all orientations within azimuth $\phi_{RX}\in~[-180^\circ, 180^\circ)$ and elevation (el.) $\theta_{RX}\in~[-30^\circ, 60^\circ]$ at each RX position using a combination of electronic beamsteering by the phased antenna array and, when the electronic beamsteering range was exceeded, mechanical movement by a 3D turntable (\emph{cf.} Fig.~\ref{subfig:turntable}). The azimuth and elevation resolution of $6^\circ$ and $30^\circ$ were chosen based on the H- and E-plane HPBW of $6^\circ$ and $36^\circ$, respectively (\emph{cf.}~Figs.~\ref{subfig:16x2patternH}--\ref{subfig:16x2patternEl}). The measurement time per single RX orientation $(\phi_{RX}, \theta_{RX})$ was 1~s. A sweep over all RX orientations for a single fixed TX orientation took 500~s. This was repeated for all TX orientations in the range $\phi_{TX}\in [-90^\circ, 84^\circ]$ with a resolution of $\Delta \phi_{TX} = 6^\circ$. The phased antenna array TX was also mounted on a 3D turntable, with the elevation angle fixed to $\theta_{TX}=-15^\circ$ based on the large E-Plane HPBW and the higher positioning on the rooftop of a parking lot. Due to the time-consuming nature of such fine-grained angular measurements, we limited the number of measurement positions to five qualitatively distinct TX-RX pairs. The overall angular coverage per position is presented in Table~\ref{tab:measurements}.

\subsubsection{Horn Antenna Transceiver}

To mechanically steer the transceiver in various orientations, the horn antenna TX and RX setups were mounted on 3D turntables, similar to that shown in Fig.~\ref{subfig:turntable}. We covered all RX azimuth angles $\phi_{RX}\in~[-180^\circ, 180^\circ)$ with a resolution $\Delta\phi_{RX}=6^\circ$ and the RX elevation angles $\theta_{RX}\in[-30^\circ, 60^\circ]$ with a resolution $\Delta \theta_{RX}=10^\circ$. On the TX side, the elevation angle was fixed to ${\theta_{TX}=-15^\circ}$, in accordance with the E-Plane HPBW of $34^\circ$ (\emph{cf.}~Fig.~\ref{subfig:FM25240_HE}). A measurement run for a fixed TX orientation took considerably longer with the horn antenna setup than with the phased antenna array setup, since all movements were achieved by mechanical movement of the 3D turntable. Therefore, we limited the range of horn antenna TX azimuth angles per measurement position ensuring that a minimum range of $54^\circ$ centered around the LOS direction was systematically covered with a $\Delta \phi_{TX}=6^\circ$ granularity. The overall angular coverage per position is presented in Table~\ref{tab:measurements}.

\vspace{-1ex}
\subsection{Post-Processing of Measurement Data}

In our analysis, we take two different perspectives on our measurement results. First, we study the link opportunities in a typical European town in Sec.~\ref{subsec:linkopportunities}, i.e. we are interested in the path loss between TX and RX for different antenna orientations. Therefore, we present the RSS of both measurement setups for a normalized nominal link budget to eliminate effects that may occur due to different antenna gains or different equivalent isotropically radiated powers (EIRPs). Second, we estimate the achievable data rate to analyze the effect of beam misalignment on the link performance, presented in Sec.~\ref{subsec:beammisanalysis}. In the following, we detail the post-processing steps taken to obtain the data presented in Sec.~\ref{subsec:linkopportunities} and Sec.~\ref{subsec:beammisanalysis} from our RSS measurements.

\subsubsection*{RSS with Normalized Nominal Link Budget}

\begin{table}[tb]
	\centering
	\caption{System parameters of the antenna setups and the wideband 5G NR system model.}
	\label{tab:systemparameters}
	\small
	\begin{tabular}{p{3cm} >{\centering\arraybackslash}p{1.3cm} >{\centering\arraybackslash}p{1.5cm} >{\centering\arraybackslash}p{1.35cm}}
		\toprule
		& Horn ant. \newline setup & Phased \newline array setup & 5G-NR model\\
		\midrule
		Main lobe gain $G$ (dBi) &  15 &  16 & -- \\
		EIRP (dBm) &  9.61 &  3.95 &  25 \\
		Sensitivity $S$ (dBm) & -91 & -104 & -74.4 \\
		\bottomrule
	\end{tabular}
\end{table}

We obtain the normalized RSS for the phased antenna array setup $RSS_A$ and the horn antenna setup $RSS_H$ from the measured RSS $RSS_{A,meas}$ and $RSS_{H,meas}$, respectively, by making equal the nominal link budget for both measurement setups, i.e.
\begin{eqnarray}
	RSS_{A} &=& RSS_{A, meas} + C_c , \\
	RSS_{H} &=& RSS_{H, meas}.
\end{eqnarray}
The constant $C_c$ is the difference between the actual link budgets of the setups, i.e.
\begin{equation}
	C_c = EIRP_{H} + G_{H} - (EIRP_{A} + G_{A} ) ,
\end{equation}
where $EIRP_{H}$ and $EIRP_{A}$ are the EIRPs of the horn antenna setup and the phased antenna array setup during the measurements, respectively, and $G_{H}$ and $G_{A}$ are the estimated main lobe antenna gains of the horn antenna and the phased antenna array, respectively. The values of these parameters are given in Table~\ref{tab:systemparameters}. We note that the validity of our results is not compromised by this operation as the sensitivity of the phased antenna array setup $S_{A,meas}$ was considerably lower than the sensitivity of the horn antenna setup $S_{H,meas}$. The sensitivity of our overall measurement system after the normalization is still limited by the horn antenna measurement setup, i.e. 
\begin{eqnarray}
 S = \mbox{max}\{S_{H,meas}, S_{A,meas} + C_c \} = -91\; \mbox{dBm},
\end{eqnarray} 
which yields a dynamic range of 115 dB. This is sufficient for our study of spatial link opportunities and beam misalignment effects on achievable data rate, as very weak signals are not relevant for stable high-speed mm-wave links and practical beamtraining. The normalized RSSs $RSS_A$ and $RSS_H$ obtained after this operation are presented in Sec.~\ref{subsec:linkopportunities}.

\subsubsection*{Estimating Achievable Data Rate}
To analyze the effect of beam misalignment on the link performance, we mapped $RSS_{A, meas}$ and $RSS_{H, meas}$ to the estimated achievable data rate $T$, assuming a fixed BS EIRP of $EIRP_{BS} = 25$~dBm, in line with existing regulations at 60~GHz \cite{EIRP_Germany}. First, the measured received signal strengths $RSS_{A, meas}$ and $RSS_{H, meas}$ were scaled according to the additional power that receivers would receive for a BS transmitting with $EIRP_{BS}$, i.e.
\begin{eqnarray}
	RSS'_{A} &=& RSS_{A, meas} + (EIRP_{BS} - EIRP_{A}) ,\\
	RSS'_{H} &=& RSS_{H, meas} + (EIRP_{BS} - EIRP_{H}). 
\end{eqnarray}
The overall sensitivity $S'$ of our overall measurement system is then given by
\begin{eqnarray}
	S' &=& \mbox{max}\{S_{H,meas} + (EIRP_{BS} - EIRP_{H}), \nonumber  \\
	     &&  S_{A,meas} +  (EIRP_{BS} - EIRP_{A}) \} \\ 
               &=& -75.61 \; \mbox{dBm}.
\end{eqnarray}
To then map $RSS'_{A}$ and $RSS'_{H}$ to the estimated, achievable data rate $T$ for a wideband system, we used the Verizon 5G-NR model which demonstrated a maximum throughput of 4~Gbps using an 800~MHz bandwidth \cite{5G-NR}. An attenuated and truncated version of the Shannon bound is used as a mapping function, similar to those used for modeling of link adaption in LTE \cite{3GPP_LTE}, i.e.
\begin{eqnarray}
\label{eq:rssdatarate}
T = \begin{cases} 0 & \mbox{for } SNR < SNR_{min}, \\
\multirow{2}{*}{$\alpha W\log_2 (1+SNR)$} & \mbox{for } SNR_{min} \le SNR \\ 
	& \mbox{and }  SNR \le  SNR_{max}, \\
T_{max} & \mbox{for } SNR > SNR_{max},
\end{cases}
\end{eqnarray}
where $W=800$~MHz denotes the bandwidth, $\alpha~=~0.75$ is the correction factor for implementation losses, $SNR_{min}=-4.5$ dB, and $SNR_{max}=~20$~dB. The modeled receiver is assumed to have a noise figure of $NF=10$~dB and implementation loss of $L_I=5$~dB such that
\begin{eqnarray}
SNR = RSS'_{A/H} - ( N + NF + L_I )
\end{eqnarray}  
where $N = 10\log_{10}(kBT)$ with Boltzmann constant $k$ is the thermal noise for assumed temperature $T=290$~K. We note that this post-processing step does not compromise the validity of our measured data since the sensitivity $S'$ is still sufficiently low to ensure detection of all signals that result in a data rate greater than zero, i.e.
\begin{eqnarray}
    SNR(RSS=S') = -5.61 \; \mbox{dB} < SNR_{min}.
\end{eqnarray} 
The estimated data rate $T$ obtained after this post-processing operation is presented in Sec.~\ref{subsec:beammisanalysis}.

\section{Measurement Results \& Analysis}
\label{sec:results}
In this section, we present the results of the measurement campaign outlined in Sec.~\ref{sec:setup}. In Sec.~\ref{subsec:linkopportunities}, we analyze available link opportunities based on measured RSS. In Sec.~\ref{subsec:beammisanalysis}, we analyze the effects of beam misalignment on the estimated achievable mm-wave data rate.

\subsection{Link Opportunities \& AoA Analysis}
\label{subsec:linkopportunities}

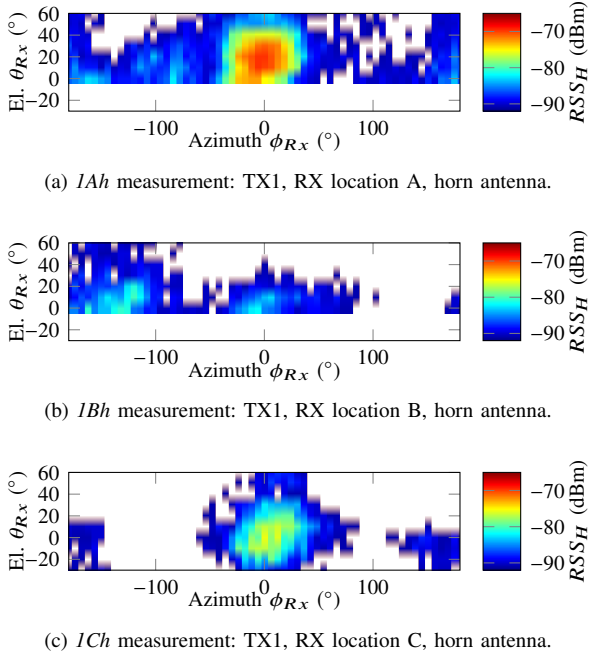
\begin{figure}[t]
	\centering
	\subfloat[\textit{1Ah} measurement: TX1, RX location A, horn antenna.\label{subfig:maxrsshorn_rxa}]			{
%
%
\begin{tikzpicture}

\begin{axis}[%
width=148pt,
height=37pt,
at={(0pt,0pt)},
scale only axis,
point meta min=-92,
point meta max=-65,
axis on top,
xmin=-180,
xmax=180,
xlabel near ticks,
xlabel shift={-8pt},
xlabel style={font={\footnotesize\color{black!100}}},
xlabel={$\text{Azimuth }\phi_{Rx}\;(^\circ\text{)}$},
ymin=-30,
ymax=60,
ylabel near ticks,
ylabel shift={-8pt},
ylabel style={font={\footnotesize\color{black!100}}},
ylabel={$\text{El. }\theta_{Rx}\;(^\circ\text{)}$},
axis background/.style={fill=white},
legend style={legend cell align=left, align=left, draw=white!15!black},
unit vector ratio={1 1},ticklabel style={font={\color{black}\footnotesize}},
colormap/jet,
colorbar,
colorbar style={ylabel near ticks, ylabel shift={-5pt},ylabel style={font={\footnotesize\color{black!100}}}, ylabel={$RSS_H$ (dBm)}}
]
\addplot [forget plot] graphics [xmin=-180.25, xmax=180.25, ymin=-5, ymax=65] {TX1_RXA_Horn_RSS_MAX-1.png};
\end{axis}
\end{tikzpicture}%
	} \\
	\subfloat[\textit{1Bh} measurement: TX1, RX location B, horn antenna.\label{subfig:maxrsshorn_rxb}]			{
%
%
\begin{tikzpicture}

\begin{axis}[%
width=148pt,
height=37pt,
at={(0pt,0pt)},
scale only axis,
point meta min=-92,
point meta max=-65,
axis on top,
xmin=-180,
xmax=180,
xlabel near ticks,
xlabel shift={-8pt},
xlabel style={font={\footnotesize\color{black!100}}},
xlabel={$\text{Azimuth }\phi_{Rx}\;(^\circ\text{)}$},
ymin=-30,
ymax=60,
ylabel near ticks,
ylabel shift={-8pt},
ylabel style={font={\footnotesize\color{black!100}}},
ylabel={$\text{El. }\theta_{Rx}\;(^\circ\text{)}$},
axis background/.style={fill=white},
legend style={legend cell align=left, align=left, draw=white!15!black},
unit vector ratio={1 1},ticklabel style={font={\color{black}\footnotesize}},
colormap/jet,
colorbar,
colorbar style={ylabel near ticks, ylabel shift={-5pt},ylabel style={font={\footnotesize\color{black!100}}}, ylabel={$RSS_H$ (dBm)}}
]
\addplot [forget plot] graphics [xmin=-180.25, xmax=180.25, ymin=-5, ymax=65] {TX1_RXB_Horn_RSS_MAX-1.png};
\end{axis}
\end{tikzpicture}%
	} \\
	\subfloat[\textit{1Ch} measurement: TX1, RX location C, horn antenna.\label{subfig:maxrsshorn_rxc}]			{
%
%
\begin{tikzpicture}

\begin{axis}[%
width=148pt,
height=37pt,
at={(0pt,0pt)},
scale only axis,
point meta min=-92,
point meta max=-65,
axis on top,
xmin=-180,
xmax=180,
xlabel near ticks,
xlabel shift={-8pt},
xlabel style={font={\footnotesize\color{black!100}}},
xlabel={$\text{Azimuth }\phi_{Rx}\;(^\circ\text{)}$},
ymin=-30,
ymax=60,
ylabel near ticks,
ylabel shift={-8pt},
ylabel style={font={\footnotesize\color{black!100}}},
ylabel={$\text{El. }\theta_{Rx}\;(^\circ\text{)}$},
axis background/.style={fill=white},
legend style={legend cell align=left, align=left, draw=white!15!black},
unit vector ratio={1 1},ticklabel style={font={\color{black}\footnotesize}},
colormap/jet,
colorbar,
colorbar style={ylabel near ticks, ylabel shift={-5pt},ylabel style={font={\footnotesize\color{black!100}}}, ylabel={$RSS_H$ (dBm)}}
]
\addplot [forget plot] graphics [xmin=-180.25, xmax=180.25, ymin=-35, ymax=65] {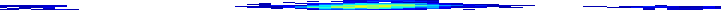};
\end{axis}
\end{tikzpicture}%
	} 
	\caption{RSS versus RX orientation for \emph{horn antenna} measurements at different TX-RX locations (assuming optimal TX orientation). \label{fig:maxrsshorn}}
\end{figure}

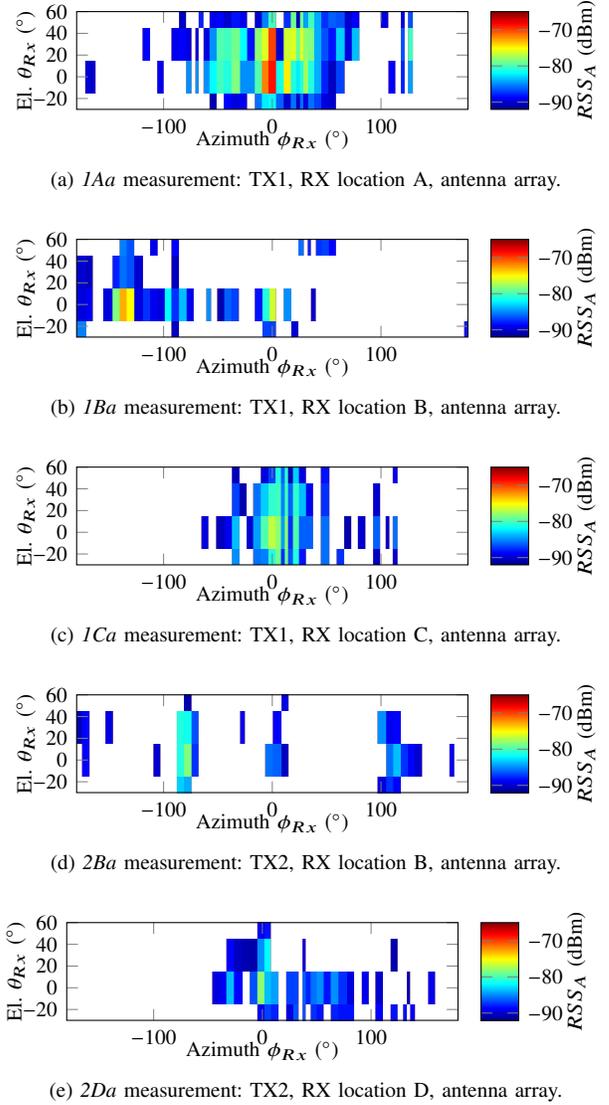
\begin{figure}[t]
	\centering
	\subfloat[\textit{1Aa} measurement: TX1, RX location A, antenna array.\label{subfig:maxrsstrx_rxa}]			{
%
%
\begin{tikzpicture}

\begin{axis}[%
width=148pt,
height=37pt,
at={(0pt,0pt)},
scale only axis,
point meta min=-92,
point meta max=-65,
axis on top,
xmin=-180,
xmax=180,
xlabel near ticks,
xlabel shift={-8pt},
xlabel style={font={\footnotesize\color{black!100}}},
xlabel={$\text{Azimuth }\phi_{Rx}\;(^\circ\text{)}$},
ymin=-30,
ymax=60,
ylabel near ticks,
ylabel shift={-8pt},
ylabel style={font={\footnotesize\color{black!100}}},
ylabel={$\text{El. }\theta_{Rx}\;(^\circ\text{)}$},
axis background/.style={fill=white},
legend style={legend cell align=left, align=left, draw=white!15!black},
unit vector ratio={1 1},ticklabel style={font={\color{black}\footnotesize}},
colormap/jet,
colorbar,
colorbar style={ylabel near ticks, ylabel shift={-5pt},ylabel style={font={\footnotesize\color{black!100}}}, ylabel={$RSS_A$ (dBm)}}
]
\addplot [forget plot] graphics [xmin=-180.25, xmax=180.25, ymin=-45, ymax=75] {TX1_RXA_TRX_RSS_MAX_corr-1.png};
\end{axis}
\end{tikzpicture}%
	}  \\
	\subfloat[\textit{1Ba} measurement: TX1, RX location B, antenna array.\label{subfig:maxrsstrx_rxb}]			{
%
%
\begin{tikzpicture}

\begin{axis}[%
width=148pt,
height=37pt,
at={(0pt,0pt)},
scale only axis,
point meta min=-92,
point meta max=-65,
axis on top,
xmin=-180,
xmax=180,
xlabel near ticks,
xlabel shift={-8pt},
xlabel style={font={\footnotesize\color{black!100}}},
xlabel={$\text{Azimuth }\phi_{Rx}\;(^\circ\text{)}$},
ymin=-30,
ymax=60,
ylabel near ticks,
ylabel shift={-8pt},
ylabel style={font={\footnotesize\color{black!100}}},
ylabel={$\text{El. }\theta_{Rx}\;(^\circ\text{)}$},
axis background/.style={fill=white},
legend style={legend cell align=left, align=left, draw=white!15!black},
unit vector ratio={1 1},ticklabel style={font={\color{black}\footnotesize}},
colormap/jet,
colorbar,
colorbar style={ylabel near ticks, ylabel shift={-5pt},ylabel style={font={\footnotesize\color{black!100}}}, ylabel={$RSS_A$ (dBm)}}
]
\addplot [forget plot] graphics [xmin=-180.25, xmax=180.25, ymin=-45, ymax=75] {TX1_RXB_TRX_RSS_MAX_corr-1.png};
\end{axis}
\end{tikzpicture}%
	} \\
	\subfloat[\textit{1Ca} measurement: TX1, RX location C, antenna array.\label{subfig:maxrsstrx_rxc}]			{
%
%
\begin{tikzpicture}

\begin{axis}[%
width=148pt,
height=37pt,
at={(0pt,0pt)},
scale only axis,
point meta min=-92,
point meta max=-65,
axis on top,
xmin=-180,
xmax=180,
xlabel near ticks,
xlabel shift={-8pt},
xlabel style={font={\footnotesize\color{black!100}}},
xlabel={$\text{Azimuth }\phi_{Rx}\;(^\circ\text{)}$},
ymin=-30,
ymax=60,
ylabel near ticks,
ylabel shift={-8pt},
ylabel style={font={\footnotesize\color{black!100}}},
ylabel={$\text{El. }\theta_{Rx}\;(^\circ\text{)}$},
axis background/.style={fill=white},
legend style={legend cell align=left, align=left, draw=white!15!black},
unit vector ratio={1 1},ticklabel style={font={\color{black}\footnotesize}},
colormap/jet,
colorbar,
colorbar style={ylabel near ticks, ylabel shift={-5pt},ylabel style={font={\footnotesize\color{black!100}}}, ylabel={$RSS_A$ (dBm)}}
]
\addplot [forget plot] graphics [xmin=-180.25, xmax=180.25, ymin=-45, ymax=75] {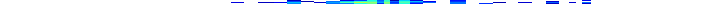};
\end{axis}
\end{tikzpicture}%
	} \\
	\subfloat[\textit{2Ba} measurement: TX2, RX location B, antenna array.\label{subfig:maxrsstrx_2rxb}]			{
%
%
\begin{tikzpicture}

\begin{axis}[%
width=148pt,
height=37pt,
at={(0pt,0pt)},
scale only axis,
point meta min=-92,
point meta max=-65,
axis on top,
xmin=-180,
xmax=180,
xlabel near ticks,
xlabel shift={-8pt},
xlabel style={font={\footnotesize\color{black!100}}},
xlabel={$\text{Azimuth }\phi_{Rx}\;(^\circ\text{)}$},
ymin=-30,
ymax=60,
ylabel near ticks,
ylabel shift={-8pt},
ylabel style={font={\footnotesize\color{black!100}}},
ylabel={$\text{El. }\theta_{Rx}\;(^\circ\text{)}$},
axis background/.style={fill=white},
legend style={legend cell align=left, align=left, draw=white!15!black},
unit vector ratio={1 1},ticklabel style={font={\color{black}\footnotesize}},
colormap/jet,
colorbar,
colorbar style={ylabel near ticks, ylabel shift={-5pt},ylabel style={font={\footnotesize\color{black!100}}}, ylabel={$RSS_A$ (dBm)}}
]
\addplot [forget plot] graphics [xmin=-180.25, xmax=180.25, ymin=-45, ymax=75] {TX2_RXB_TRX_RSS_MAX_corr-1.png};
\end{axis}
\end{tikzpicture}%
	} \\
	\subfloat[\textit{2Da} measurement: TX2, RX location D, antenna array.\label{subfig:maxrsstrx_2rxd}]			{
%
%
\hspace*{-11pt}
\begin{tikzpicture}

\begin{axis}[%
width=148pt,
height=37pt,
at={(0pt,0pt)},
scale only axis,
point meta min=-92,
point meta max=-65,
axis on top,
xmin=-180,
xmax=180,
xlabel near ticks,
xlabel shift={-8pt},
xlabel style={font={\footnotesize\color{black!100}}},
xlabel={$\text{Azimuth }\phi_{Rx}\;(^\circ\text{)}$},
ymin=-30,
ymax=60,
ylabel near ticks,
ylabel shift={-8pt},
ylabel style={font={\footnotesize\color{black!100}}},
ylabel={$\text{El. }\theta_{Rx}\;(^\circ\text{)}$},
axis background/.style={fill=white},
legend style={legend cell align=left, align=left, draw=white!15!black},
unit vector ratio={1 1},ticklabel style={font={\color{black}\footnotesize}},
colormap/jet,
colorbar,
colorbar style={ylabel near ticks, ylabel shift={-5pt},ylabel style={font={\footnotesize\color{black!100}}}, ylabel={$RSS_A$ (dBm)}}
]
\addplot [forget plot] graphics [xmin=-180.25, xmax=180.25, ymin=-45, ymax=75] {TX2_RXD_TRX_RSS_MAX_corr-1.png};
\end{axis}
\end{tikzpicture}%
	}%
	\caption{RSS  versus RX orientation for \emph{phased antenna array} measurements at different TX-RX locations (assuming optimal TX orientation). \label{fig:maxrssarray}}
\end{figure}

\begin{figure*}[t]
	\centering
	\setlength{\fboxsep}{0pt}%
	\subfloat[\emph{1A} measurement.\label{subfig:aoa_tx1rxa}]{
		\fbox{
			\includegraphics[width=0.22\linewidth]{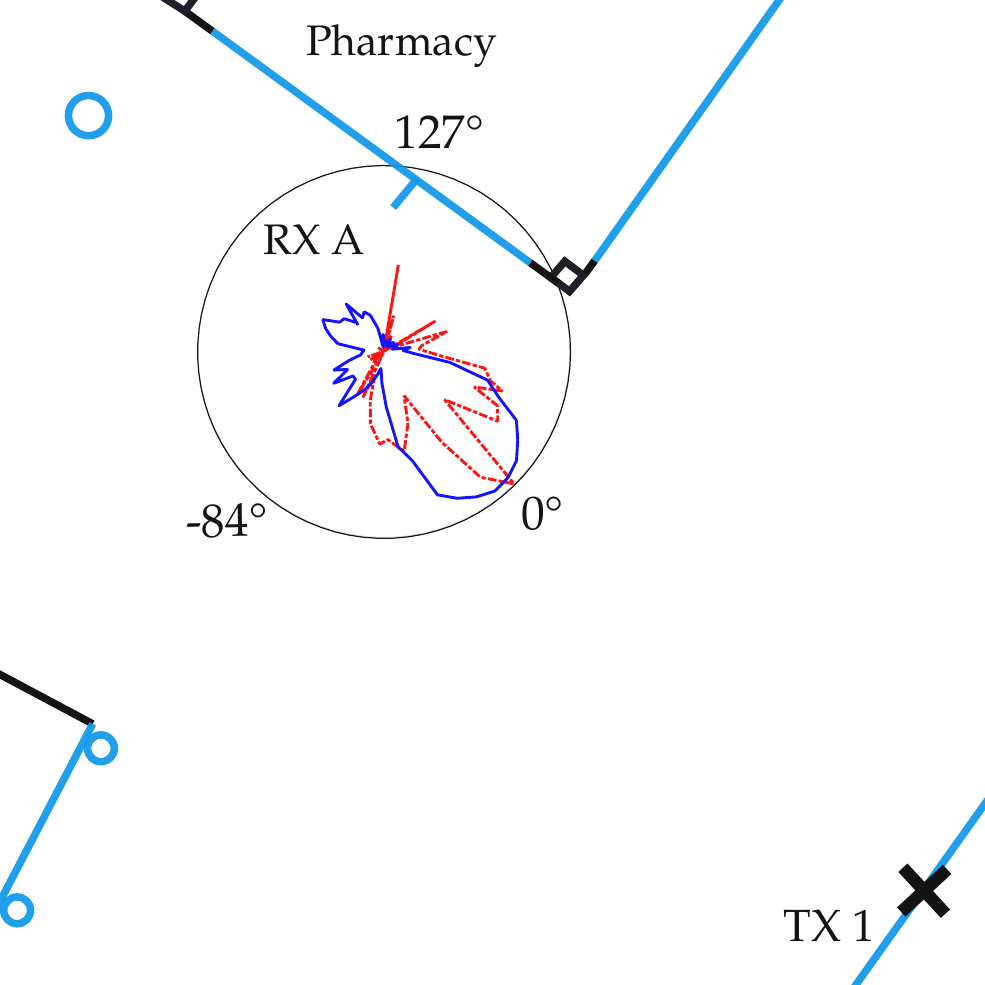}
		}
	} 
	\subfloat[\emph{1B} measurement.\label{subfig:aoa_tx1rxb}]{
		\fbox{
			\includegraphics[width=0.22\linewidth]{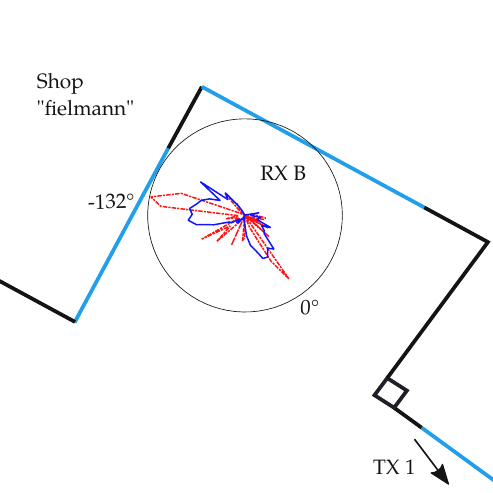}
		}
	} 
	\subfloat[\emph{1C} measurement.\label{subfig:aoa_tx1rxc}]{
		\fbox{
			\includegraphics[width=0.22\linewidth]{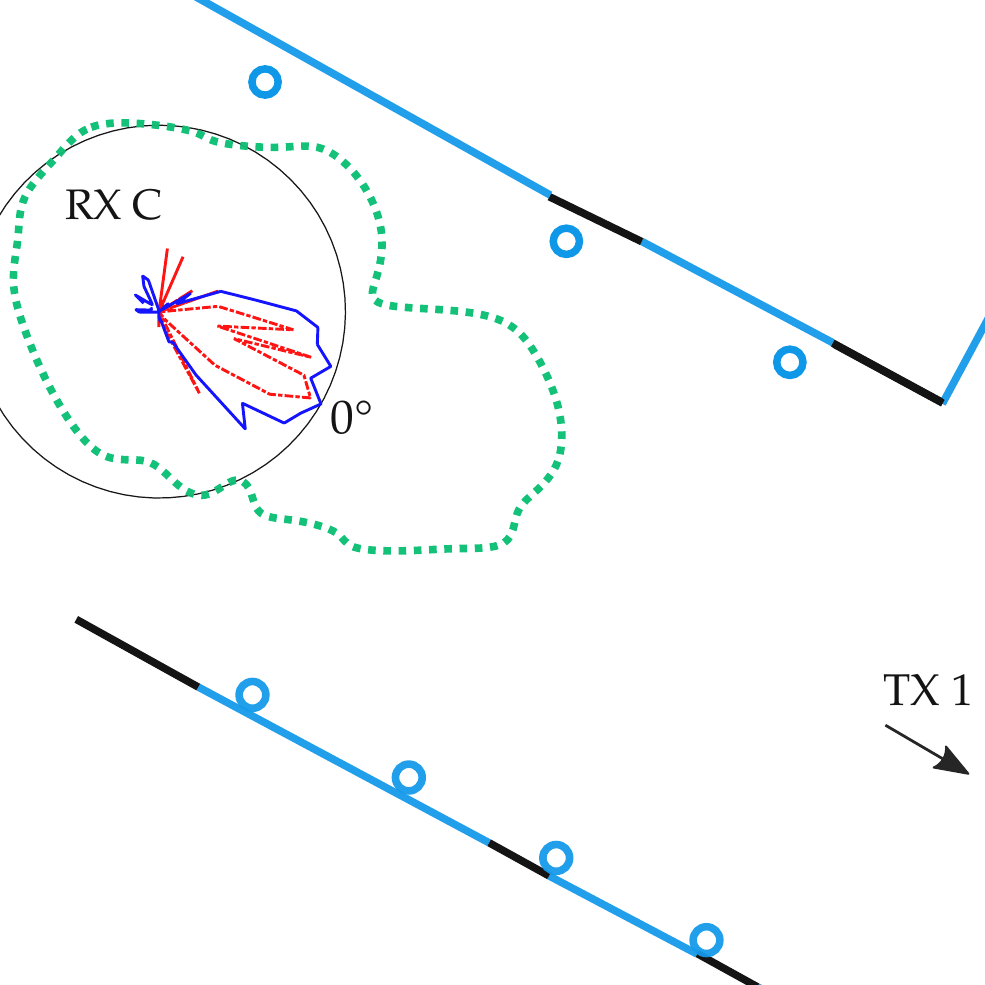}
		}
	} \\
	\subfloat[\emph{2B} measurement.\label{subfig:aoa_tx2rxb}]{
		\fbox{
			\includegraphics[width=0.22\linewidth]{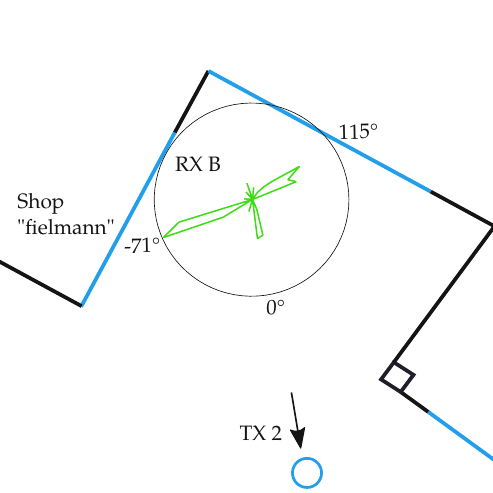}
		}
	} 
	\subfloat[\emph{2D} measurement.\label{subfig:aoa_tx2rxd}]{
		\fbox{
			\includegraphics[width=0.22\linewidth]{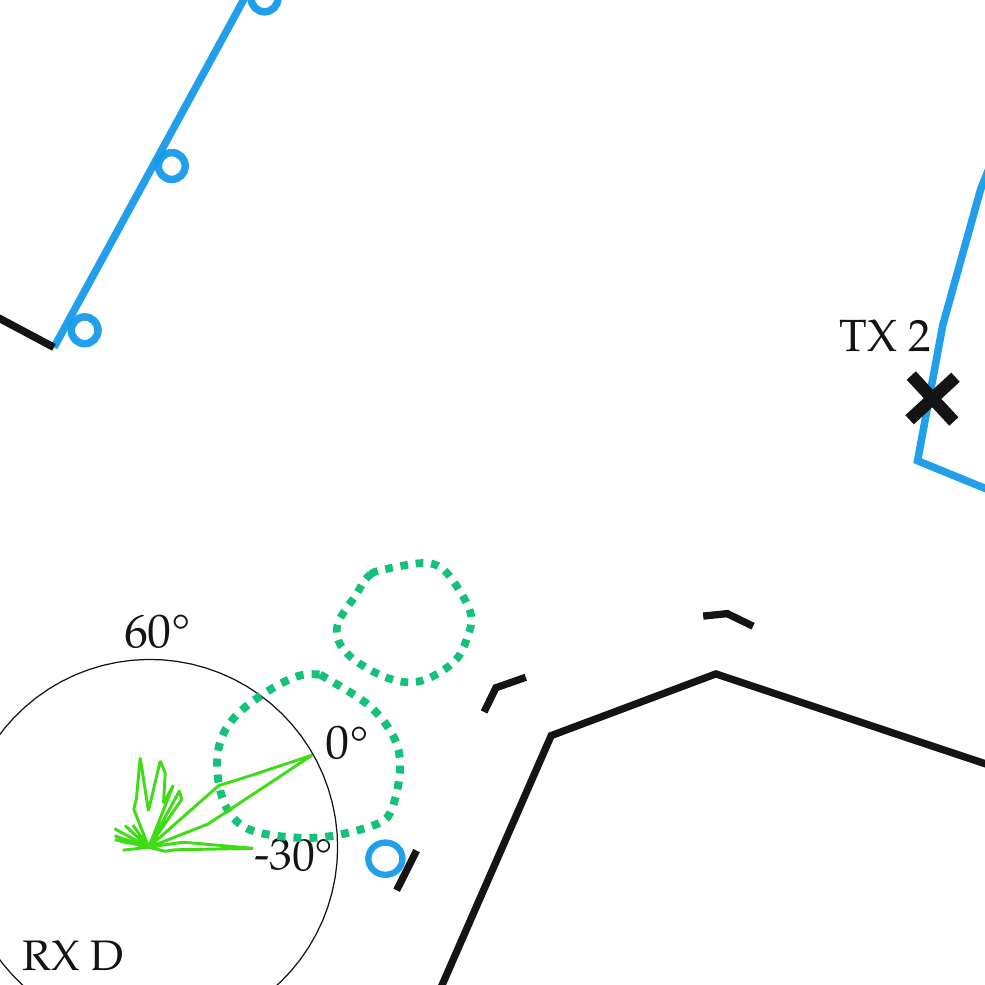}
		}
	}
	\subfloat[Legend.\label{subfig:aoa_legend}]{
		\fbox{
			\includegraphics[width=0.22\linewidth]{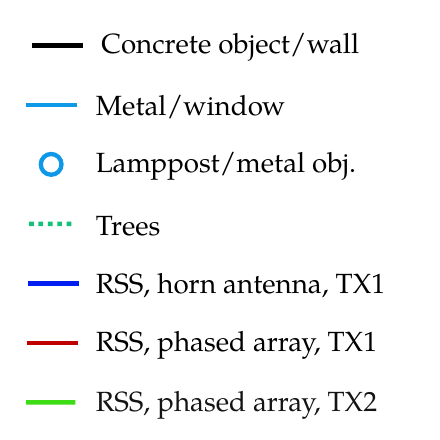}
		}
	}
	\caption{Illustrated RSS over azimuth angles per TX-RX pair. The RSS polar plots are normalized to the maximum RSS per RX location to focus on the AoA analysis.\label{fig:aoa}}
\end{figure*}

\begin{figure}[t]
	\centering
	\subfloat[Signal paths from transmitter 1.\label{subfig:paths_tx1}]{
		\includegraphics[width=0.3\textwidth]{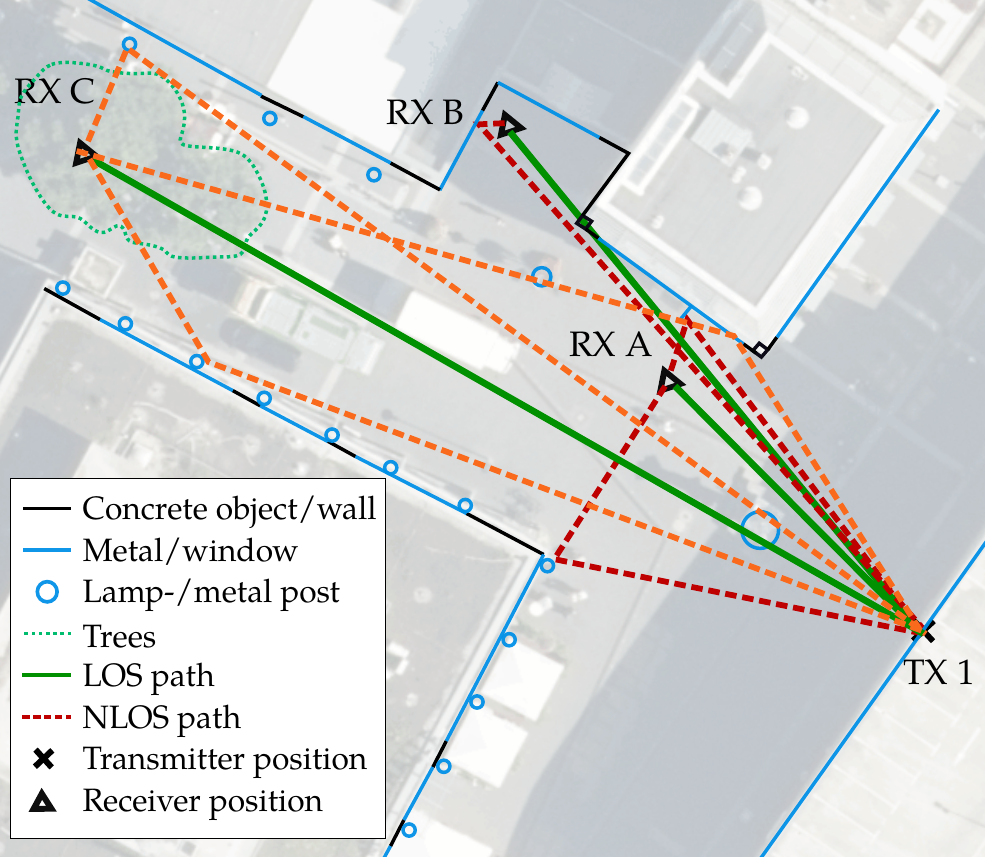}
	}
	\\
	\subfloat[Signal paths from transmitter 2.\label{subfig:paths_tx2}]{
		\includegraphics[width=0.3\textwidth, height=0.3\textheight]{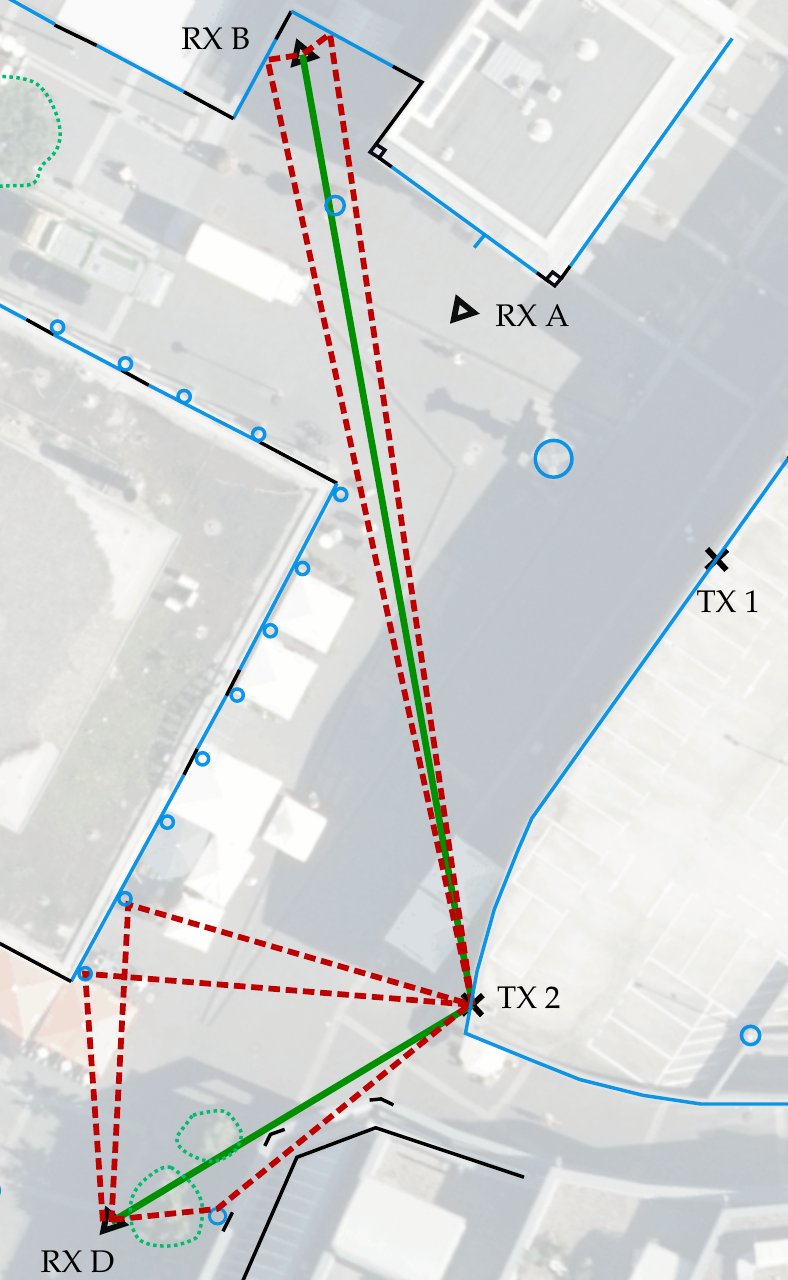}
	}
	\caption{Maps showing the LOS/NLOS path traces from the different TX and RX positions. The materials of buildings and other objects are also indicated.\label{fig:paths}}
\end{figure}

In the following, we study significant link opportunities found during our measurement campaign, trace the corresponding physical propagation paths on a map and analyze the observed differences between the phased antenna array and horn antenna data. We present our results as follows. Figs.~\ref{fig:maxrsshorn}-\ref{fig:maxrssarray} present the RSS versus RX orientations in a heatmap from RX perspective, i.e. the RSS per RX orientation $(\phi_{RX}, \theta_{RX})$. This allows us to investigate the structure of received multipath clusters. We assume for each RX orientation $(\phi_{RX}, \theta_{RX})$ the best corresponding TX orientation $\phi_{TX}$ found during the measurements to show the full set of multipath clusters available at each RX location. Fig.~\ref{fig:aoa} illustrates the same underlying data as in Figs.~\ref{fig:maxrsshorn}-\ref{fig:maxrssarray}, showing polar plots of the RSS on a schematic map. Finally, in Fig.~\ref{fig:paths} we trace the corresponding independent physical propagation paths between TX and RX based on the AoA, AoD, and potential reflectors. In the following, our results are presented based on Figs.~\ref{fig:maxrsshorn}-\ref{fig:paths} location-by-location.

We start our analysis with a clear LOS scenario, i.e. at RX location~A. A strong cluster around the LOS, i.e. at $\phi_{RX}=0^\circ$, is observed for both measurement setups in the heatmaps in Figs.~\ref{subfig:maxrsshorn_rxa}--\ref{subfig:maxrsstrx_rxa}. Nonetheless, we observe the distinct effect of the different antenna patterns on the cluster appearance. While the horn antenna measurement in Fig.~\ref{subfig:maxrsshorn_rxa} exhibits a smooth cluster, the same cluster appears irregular for the phased antenna array measurement in Fig.~\ref{subfig:maxrsstrx_rxa}\footnote{In Fig.~\ref{fig:maxrssarray}, we note that the measurement points are not spaced equidistantly over azimuth angles $\phi_{RX}$. This is due to the imperfect beam steering with the manufacturer-supplied beambook, i.e. we observed an offset between the beam steering angle that we set and the actual angle of the main lobe seen at the over-the-air interface. We measured the offset for all beamsteering angles and correct for these offsets during the post-processing of the results.}. Studying the AoA polar plot in Fig.~\ref{subfig:aoa_tx1rxa}, we recognize that the non-ideal, phased antenna array radiation pattern (\emph{cf.} Fig.~\ref{subfig:16x2patternH}) makes it much more difficult to identify true independent propagation paths. We can identify that the first peak around $\phi_{RX}=0^\circ$ points in the LOS direction and, using AoD information (not shown for brevity), that the peaks at $\phi_{RX}=-84^\circ$ and $\phi_{RX}=127^\circ$ correspond to independent secondary NLOS paths as shown in Fig.~\ref{subfig:paths_tx1}. By contrast, the peaks around $\phi_{RX}=-30^\circ$ and $\phi_{RX}=30^\circ$ are due to sidelobes that point into the LOS direction, thereby falsely indicating independent propagation paths. This is clear when looking at the map in Fig.~\ref{subfig:aoa_tx1rxa}, as there are no reflectors at $\phi_{RX}=-30^\circ$ and $\phi_{RX}=30^\circ$. This may cause problems to beamsteering algorithms that do not have information about the environment and the AoD. For instance, an algorithm may generate a list of orientations supposedly corresponding to viable independent propagation paths during initial beam training. Then in the case of detected blockage on the primary path, it may switch to a falsely indicated propagation path, i.e. it switches to a sidelobe over the same path. Consequently, it would fail to overcome the link blockage.


We next consider the NLOS RX location~B. Comparing the heatmaps and AoA polar plots of the horn antenna and phased antenna array measurements in Figs.~\ref{subfig:maxrsshorn_rxb}--\ref{subfig:maxrsstrx_rxb} and Fig.~\ref{subfig:aoa_tx1rxb}, respectively, we note again the significant differences in the cluster appearance. Combining knowledge of the AoA, the environment, and the steered phased antenna array radiation pattern, we can trace two physical propagation paths as shown in Fig.~\ref{subfig:paths_tx1}. We point out that AoD information is not helpful in this scenario to distinguish between a true independent propagation path and a sidelobe-induced high RSS as the AoD is the same for both the LOS and NLOS path (\emph{cf.}~Fig.~\ref{subfig:paths_tx1}). Overall, we draw two conclusions based on these observations. First, the observation of having multiple link opportunities is promising for mm-wave system designers as it suggests that mm-wave coverage can also be provided for urban scenarios with a partly blocked LOS. Second, our results suggest that even knowing both AoA and AoD may not be sufficient to reliably identify true independent propagation paths. Instead, we used both environment information and knowledge of the real steered phased antenna array radiation pattern to identify the independent propagation paths. In practice, future commercial mobile mm-wave systems may not have such perfect, \emph{a priori} information and consequently, other strategies will be needed to identify true independent propagation paths. For instance, in scenarios with high link budget, using a nearly omni-directional radiation pattern at the receiver to record the power delay profile may allow extracting the number of independent propagation paths based on time of arrival~\cite{weiler_measuring_2014}. Alternatively, environment-awareness will be essential for robust mm-wave beamsteering.

Let us now consider the LOS scenario for RX location~C. Despite the increased TX--RX distance as compared to RX location~A and the fact that RX C is positioned in an 18~m wide street canyon, the heatmaps in Figs.~\ref{subfig:maxrsshorn_rxc}--\ref{subfig:maxrsstrx_rxc} strongly resemble the heatmaps obtained for RX location~A (\emph{cf.}~Fig.~\ref{subfig:maxrsshorn_rxa} and Fig.~\ref{subfig:maxrsstrx_rxa}, respectively). We combine the knowledge of the AoA shown in Fig.~\ref{subfig:aoa_tx1rxc}, AoD (not shown for brevity) and the knowledge of the environment to trace the independent NLOS paths shown in the map in Fig.~\ref{subfig:paths_tx1}. However, we emphasize that these are difficult to trace solely based on the AoA polar plot in Fig.~\ref{subfig:aoa_tx1rxc} due to the irregular, phased antenna array radiation pattern. For instance, the AoA at RX location~C (\emph{cf.}~Fig.~\ref{subfig:aoa_tx1rxc}) looks similar to the AoA at RX location~A (\emph{cf.}~Fig.~\ref{subfig:aoa_tx1rxa}). Yet, at RX location~A the high RSS at angle $\phi_{RX}=30^\circ$ is caused by a sidelobe whereas it corresponds to a true independent propagation path at RX location~C. This underlines the difficulty of identifying viable independent physical propagation paths by one-sided RSS measurements.

We now study measurements that were conducted with the TX at location TX~2. Fig.~\ref{subfig:maxrsstrx_2rxb} shows the heatmap for RX location~B, where we note several multipath clusters. Fig.~\ref{subfig:aoa_tx2rxb} shows the AoA polar plot, where we can see the peaks around $\phi_{RX}=-71^\circ$, $\phi_{RX}=115^\circ$, and $\phi_{RX}=0^\circ$. All three orientations correspond to true independent propagation paths as shown in the map in Fig.~\ref{subfig:paths_tx2}. The independent propagation paths were traced and distinguished from sidelobe-induced high RSS using information about the environment and the steered RX phased antenna array radiation pattern (not shown for brevity). We emphasize that AoD information cannot help to distinguish true independent propagation paths from sidelobes, as previously described for measurement \emph{1Ba}. Overall, these results support the previous observations that multiple mm-wave link opportunities can be found even in distant and NLOS scenarios in the given environment, but identifying independent propagation paths requires combined knowledge from different sources, i.e. environmental awareness, beyond pure RSS measurements. 

Finally, we show in Fig.~\ref{subfig:maxrsstrx_2rxd} the heatmap of the measurement results from RX location~D where the LOS path from the transmitter in location TX~2 was obstructed by the foliage of a tree. Nevertheless, the AoA polar plot in Fig.~\ref{subfig:aoa_tx2rxd} shows a peak towards the TX and we can identify additional NLOS paths (\emph{cf.} Fig.~\ref{subfig:paths_tx2}) based on the AoA polar plot in Fig.~\ref{subfig:aoa_tx2rxd}, AoD (not shown here for brevity), and knowledge of the environment. 

Overall, our measurements show 2--4 multipath clusters per receiver position. Reflections were mainly caused by metal or glass on facades and, somewhat surprisingly, by lampposts, enabling potential NLOS links in multiple scenarios. We observed significant differences between phased antenna array and horn antenna measurements and noted that determining independent propagation paths purely based on one-sided phased antenna array RSS measurements was not a reliable approach. Instead, knowledge of AoA, AoD, and the environment was needed to identify independent propagation paths. Additionally, due to small-scale movements of a user or non-ideal beam training, the transmitter and the receiver may not be perfectly aligned. In Sec.~\ref{subsec:beammisanalysis} we therefore present an analysis of the effects of beam misalignment at transmitter and receiver on achievable data rates of the mm-wave link opportunities revealed by our measurements. 
 
\subsection{Analysis of Beam Misalignment Effects}
\label{subsec:beammisanalysis}

\begin{figure}[t]
	\centering
	\subfloat[Scenario \textit{1Ah}: TX1, RX location~A, horn antenna.\label{subfig:estratehorn_rxa}]			{
%
%
\begin{tikzpicture}

\begin{axis}[%
width=148pt,
height=37pt,
at={(0pt,0pt)},
scale only axis,
point meta min=0.263,
point meta max=4,
axis on top,
xmin=-180,
xmax=180,
xlabel near ticks,
xlabel shift={-8pt},
xlabel style={font={\footnotesize\color{black!100}}},
xlabel={$\text{Azimuth }\phi_{Rx}\;(^\circ\text{)}$},
ymin=-30,
ymax=60,
ylabel near ticks,
ylabel shift={-8pt},
ylabel style={font={\footnotesize\color{black!100}}},
ylabel={$\text{Elevation }\theta_{Rx}\;(^\circ\text{)}$},
axis background/.style={fill=white},
legend style={legend cell align=left, align=left, draw=white!15!black},
unit vector ratio={1 1},ticklabel style={font={\color{black}\footnotesize}},
colormap/jet,
colorbar,
colorbar style={ylabel near ticks, ylabel shift={-5pt},ylabel style={font={\footnotesize\color{black!100}}}, ylabel={$T$ (Gbps)}}
]
\addplot [forget plot] graphics [xmin=-183, xmax=183, ymin=-5, ymax=65] {TX1_RXA_Horn_DATARATE-1.png};
\end{axis}
\end{tikzpicture}%
	} \\
	\subfloat[Scenario \textit{1Aa}: TX1, RX location~A, phased antenna array.\label{subfig:estratetrx_rxa}]			{
%
%
\begin{tikzpicture}

\begin{axis}[%
width=148pt,
height=37pt,
at={(0pt,0pt)},
scale only axis,
point meta min=0.263,
point meta max=4,
axis on top,
xmin=-180,
xmax=180,
xlabel near ticks,
xlabel shift={-8pt},
xlabel style={font={\footnotesize\color{black!100}}},
xlabel={$\text{Azimuth }\phi_{Rx}\;(^\circ\text{)}$},
ymin=-30,
ymax=60,
ylabel near ticks,
ylabel shift={-8pt},
ylabel style={font={\footnotesize\color{black!100}}},
ylabel={$\text{Elevation }\theta_{Rx}\;(^\circ\text{)}$},
axis background/.style={fill=white},
legend style={legend cell align=left, align=left, draw=white!15!black},
unit vector ratio={1 1},ticklabel style={font={\color{black}\footnotesize}},
colormap/jet,
colorbar,
colorbar style={ylabel near ticks, ylabel shift={-5pt},ylabel style={font={\footnotesize\color{black!100}}}, ylabel={$T$ (Gbps)}}
]
\addplot [forget plot] graphics [xmin=-180.25, xmax=180.25, ymin=-45, ymax=75] {TX1_RXA_TRX_DATARATE_corr-1.png};
\end{axis}
\end{tikzpicture}%
	}  
	\caption{Estimated data rate versus RX orientation for scenario \emph{1A} (assuming optimal TX orientation). \label{fig:estdatarate}}
\end{figure}
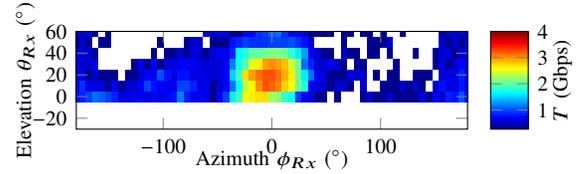
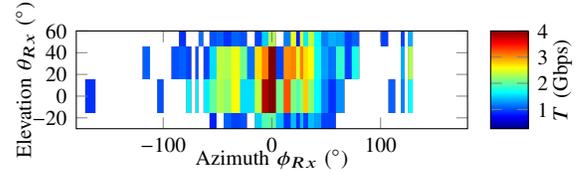

In this section we study the effect of beam misalignment on the link performance in outdoor mm-wave networks, using the estimated achievable data rate as calculated from (\ref{eq:rssdatarate}) as the performance metric. We emphasize that considering the achievable data rate rather than solely RSS measurements (as in Sec.~\ref{subsec:linkopportunities}) allows us to better assess the networking consequences of misaligned beams, especially with regard to higher system layers that need to cope with the resulting changes in the PHY data rate. To illustrate this, in Fig.~\ref{fig:estdatarate} we show the heatmaps of estimated data rate versus RX orientation for RX location~A (\emph{cf.}~Fig.~\ref{subfig:maxrsshorn_rxa} and Fig.~\ref{subfig:maxrsstrx_rxa} for the corresponding RSS heatmaps). From Fig.~\ref{subfig:estratetrx_rxa}, it is evident that a phased antenna array beam misalignment of a few degrees in the azimuth can result in a dramatic loss in data rate.

\begin{figure}[t]
	\centering
	\subfloat[BS and/or UE beam misaligned due to imperfect beam training.\label{subfig:scenariotxmisnomove}]		{
		\includegraphics[width=0.19\textwidth]{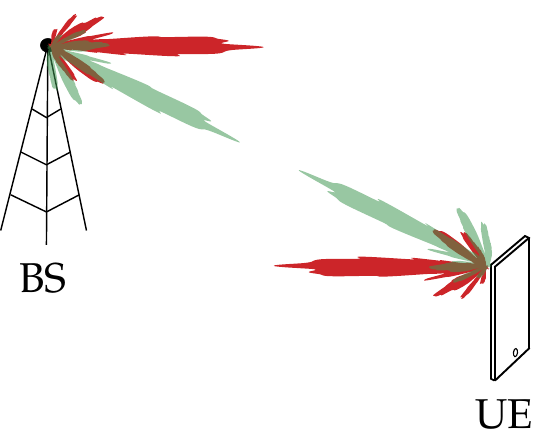}
	} 
	 \hspace{2ex}
	\subfloat[UE beam misaligned after rotation.\label{subfig:scenariorxmis}]			{
	  	\includegraphics[width=0.19\textwidth]{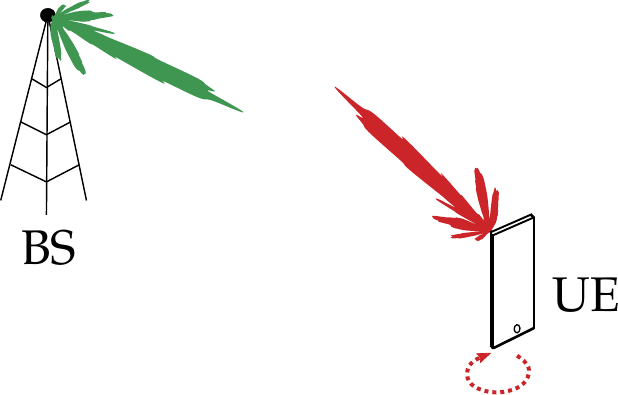}
	} \\ \vspace{2ex}
	\subfloat[BS and UE beams misaligned after small-scale lateral movement $d_{move}$ of UE.\label{subfig:scenariotxrxmis}]			{
		\includegraphics[width=0.19\textwidth]{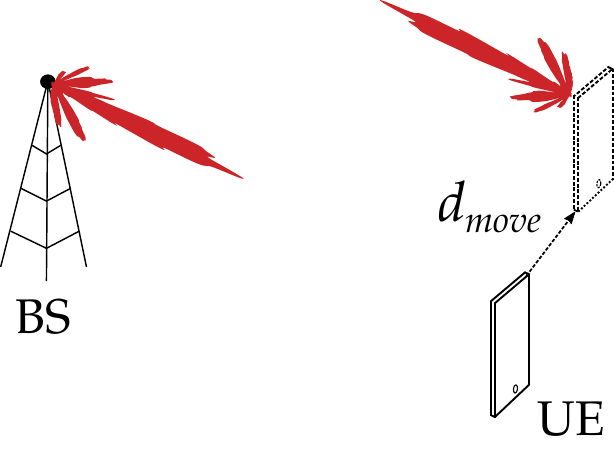}
	} \hspace{2ex}
	\subfloat[BS beam misaligned after small-scale lateral movement of UE; UE locally adjusted its beam to new optimal orientation.\label{subfig:scenariotxmis}]			{
		\includegraphics[width=0.19\textwidth]{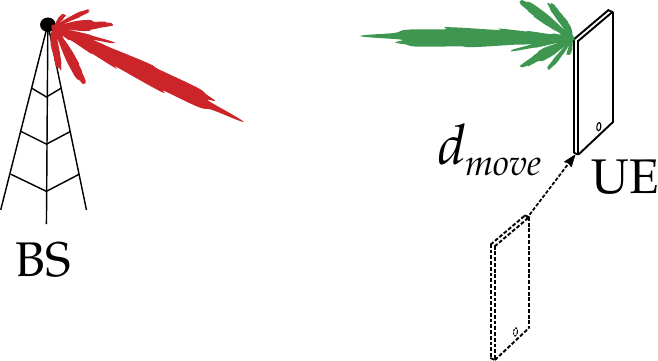}
	}   
	\caption{Scenarios illustrating potential beam misalignments at BS and UE. \label{fig:beammisscenarios}}
\end{figure}

In real mm-wave outdoor deployments beam misalignments can occur for several reasons, as illustrated in Fig.~\ref{fig:beammisscenarios}, where we consider, without loss of generality, the downlink. After successful initial beam training, the BS and UE are ideally perfectly aligned. However, imperfect beam training due to e.g. stale beam training data or ambiguous information about independent propagation paths (as described in Sec.~\ref{subsec:linkopportunities}) may result in beam misalignments as shown in Fig.~\ref{subfig:scenariotxmisnomove}. For instance, a device rotation may cause a UE beam misalignment (\emph{cf.} Fig.~\ref{subfig:scenariorxmis}) and a small-scale lateral movement of the UE out of the BS's beam may cause a misalignment at both BS and UE (\emph{cf.} Fig.~\ref{subfig:scenariotxrxmis}). The UE may readjust its beam orientation to mitigate the effects of the BS misalignment (\emph{cf.} Fig.~\ref{subfig:scenariotxmis}). In the following, we analyze how these beam misalignment scenarios affect the estimated achievable data rate based on our measured RSS data. 

\subsubsection{BS Beam Misalignment Analysis}
\label{subsec:txmisalignment}

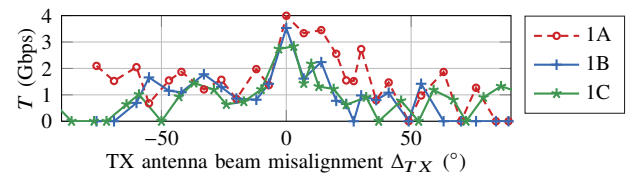
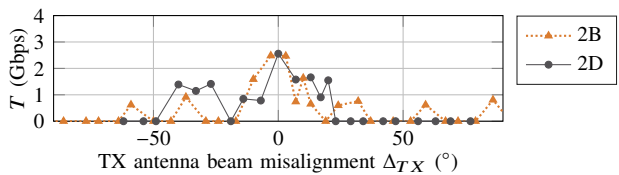
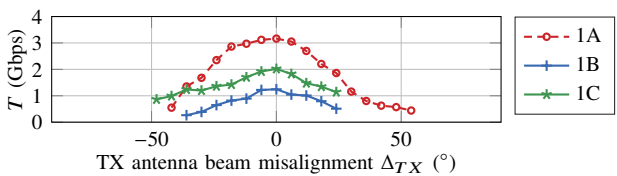
\begin{figure}
	\centering
	\subfloat[Phased antenna array results for TX1.\label{subfig:txmisalign_trx_tx1}]			{
%
%
\begin{tikzpicture}

\begin{axis}[%
width=170pt,
height=40pt,
at={(0pt,0pt)},
scale only axis,
xmin=-90,
xmax=90,
xlabel near ticks,
xlabel shift={-4pt},
xlabel style={font={\footnotesize\color{black!100}}},
xlabel={$\text{TX antenna beam misalignment } \Delta_{TX} \text{ (}^\circ\text{)}$},
ymin=0,
ymax=4,
ylabel near ticks,
ylabel shift={-5pt},
ylabel style={font={\footnotesize\color{black!100}}},
ylabel={$T$ (Gbps)},
axis background/.style={fill=white},
xmajorgrids,
ymajorgrids,
ticklabel style={font={\color{black}\footnotesize}},
legend style={legend cell align=left, align=left, draw=white!15!black},
ticklabel style={font={\color{black}\footnotesize}},legend style={font={\color{black}\footnotesize}},legend pos=outer north east,
]
\addplot [color=red, thick, densely dashed, mark=o, mark options={solid, red}, mark size=1.2pt]
table[row sep=crcr]{%
-76	2.09665158218931\\
-69	1.5265801057116\\
-60	2.04823878277648\\
-55	0.681570038915382\\
-47	1.54491970997386\\
-42	1.87047380572981\\
-33	1.20560338330928\\
-26	1.57069067880686\\
-20	0.844999920627215\\
-12	1.97428987088665\\
-7	1.36370823351449\\ 
0	3.99492688965108\\
7	3.33357525867297\\ 
14	3.45086371211678\\
20	2.55418922698426\\
24	1.54717150907041\\
27	1.51985854458017\\
30	2.73347189459133\\
36	0.798636230625633\\
41	1.46882390218231\\
49	0\\
54	0.974123053376632\\
63	1.86598290871839\\
70	0\\
76	1.26736811125095\\
84	0\\
89	0\\
96	0\\
103	0\\
110	0.72863110424026\\
};
\addlegendentry{1A}

\addplot [color=blue, thick, mark=+, mark options={solid, blue}, mark size=2pt]
  table[row sep=crcr]{%
-76	0\\
-69	0\\
-60	0.696561158385903\\
-55	1.65967218871809\\
-47	1.15556406044262\\
-42	1.09827567621591\\
-33	1.7791506060443\\
-26	1.29526854159605\\
-20	0.859532663976685\\
-12	0.814508417622997\\
-7	1.41931601648163\\ 
0	3.53106559421553\\
7	1.61646444098858\\ 
14	2.23942728540553\\
20	0.774754950469898\\
24	0.653884808806985\\
27	0\\
30	0.976599146274777\\
36	0.802298927839902\\
41	1.08183836551247\\
49	0\\
54	1.41939200323506\\
63	0\\
70	0\\
76	0\\
84	0\\
89	0\\
96	0\\
103	0\\
110	1.10965422387506\\
};
\addlegendentry{1B}

\addplot [color=green, solid, thick, mark=star, mark options={solid, green}, mark size=2pt]
  table[row sep=crcr]{%
-93	0.682840392519574\\
-86	0\\
-77	0\\
-72	0\\
-64	0.637803262498879\\
-59	1.01398850693723\\
-50	0\\
-43	0.928367741331673\\
-37	1.44959058110542\\
-29	1.19380636986401\\
-24	0.636283860938007\\
-17	0.743229258750359\\
-10	1.20486742922422\\ 
-3	2.73616131716494\\
3	2.83519813319267\\
7	1.4325248874025\\ 
10	2.17203644941714\\
13	1.31458118331678\\
19	1.22091403245407\\
24	0.623098825947234\\
32	0.908931007369732\\
37	0\\
46	0.776203700985897\\
53	0\\
59	1.18487686186165\\
67	0.807256764168841\\
72	0\\
79	0.903807676929128\\
86	1.32578645846188\\
93	1.10859535439676\\
};
\addlegendentry{1C}

\end{axis}
\end{tikzpicture}%
	} \\
	\subfloat[Phased antenna array results for TX2.\label{subfig:txmisalign_trx_tx2}]			{
%
%
\begin{tikzpicture}

\begin{axis}[%
width=170pt,
height=40pt,
at={(0pt,0pt)},
scale only axis,
xmin=-90,
xmax=90,
xlabel near ticks,
xlabel shift={-4pt},
xlabel style={font={\footnotesize\color{black!100}}},
xlabel={$\text{TX antenna beam misalignment } \Delta_{TX} \text{ (}^\circ\text{)}$},
ymin=0,
ymax=4,
ylabel near ticks,
ylabel shift={-5pt},
ylabel style={font={\footnotesize\color{black!100}}},
ylabel={$T$ (Gbps)},
axis background/.style={fill=white},
xmajorgrids,
ymajorgrids,
ticklabel style={font={\color{black}\footnotesize}},
legend style={legend cell align=left, align=left, draw=white!15!black},
ticklabel style={font={\color{black}\footnotesize}},legend style={font={\color{black}\footnotesize}},legend pos=outer north east,
]

\addplot [color=orange, densely dotted, thick, mark=triangle*, mark options={solid, orange}, mark size=1.2pt]
  table[row sep=crcr]{%
-93	0\\
-86	0\\
-77	0\\
-72	0\\
-64	0\\
-59	0.62168203185241\\
-50	0\\
-43	0\\
-37	0.913735555052071\\
-29	0\\
-24	0\\
-17	0\\
-10	1.58829149703091\\ 
-3	2.4807296796608\\
3	2.47036361204881\\
7	0.740051789433112\\ 
10	1.63482708664567\\
13	0.654637298448901\\
19	0\\
24	0.59884978749276\\
32	0.758694944674825\\
37	0\\
46	0\\
53	0\\
59	0.618034443224929\\
67	0\\
72	0\\
79	0\\
86	0.806913741533209\\
93	0\\
};
\addlegendentry{2B}

\addplot [color=grey, mark=*, mark options={solid, grey},mark size=1.3pt]
  table[row sep=crcr]{%
-62	0\\
-54	0\\
-49	0\\
-40	1.39003075056137\\
-33	1.14783624834241\\
-27	1.41294621346079\\
-19	0\\
-14	0.846490566529334\\
-7	0.783300478900245\\ 
0	2.55609018478179\\
7	1.57771636945249\\ 
13	1.66145215597833\\
17	0.90567804631877\\
20	1.5488810807461\\
23	0\\
29	0\\
34	0\\
42	0\\
47	0\\
56	0\\
63	0\\
69	0\\
77	0\\
};
\addlegendentry{2D}

\end{axis}
\end{tikzpicture}%
	} \\
	\subfloat[Horn antenna results.\label{subfig:txmisalign_horn}]			
	{
%
%
\hspace*{-11pt}
\begin{tikzpicture}

\begin{axis}[%
width=170pt,
height=40pt,
at={(0pt,0pt)},
scale only axis,
xmin=-90,
xmax=90,
xlabel near ticks,
xlabel shift={-4pt},
xlabel style={font={\footnotesize\color{black!100}}},
xlabel={$\text{TX antenna beam misalignment } \Delta_{TX} \text{ (}^\circ\text{)}$},
ymin=0,
ymax=4,
ylabel near ticks,
ylabel shift={-5pt},
ylabel style={font={\footnotesize\color{black!100}}},
ylabel={$T$ (Gbps)},
axis background/.style={fill=white},
xmajorgrids,
ymajorgrids,
ticklabel style={font={\color{black}\footnotesize}},
legend style={legend cell align=left, align=left, draw=white!15!black},
ticklabel style={font={\color{black}\footnotesize}},legend style={font={\color{black}\footnotesize}},legend pos=outer north east,
]
\addplot [color=red, thick, densely dashed, mark=o, mark options={solid, red}, mark size=1.2pt]
table[row sep=crcr]{%
-42	0.549690796107395\\
-36	1.35243090950571\\
-30	1.67977199019034\\
-24	2.35006594650216\\
-18	2.86258358522094\\
-12	2.97090030777084\\
-6	3.11782601620735\\
0	3.16409048346371\\
6	3.05327120626746\\
12	2.69663688100567\\
18	2.20434818030543\\
24	1.85982516569369\\
30	1.15857999320539\\
36	0.802545185179519\\
42	0.625557109291752\\
48	0.57062238514504\\
54	0.440648899603541\\
};
\addlegendentry{1A}

\addplot [color=blue, thick, mark=+, mark options={solid, blue}, mark size=2pt]
  table[row sep=crcr]{%
-36	0.265571390084823\\
-30	0.388378965802375\\
-24	0.647813639241614\\
-18	0.817441589103034\\
-12	0.893187429759893\\
-6	1.22152222314738\\
0	1.24484286090418\\
6	1.0452576464996\\
12	1.01009408452948\\
18	0.79363492654007\\
24	0.511171113700814\\
};
\addlegendentry{1B}

\addplot [color=green, solid, thick, mark=star, mark options={solid, green}, mark size=2pt]
  table[row sep=crcr]{%
-48	0.862734456074008\\
-42	0.989879773798098\\
-36	1.23734751954287\\
-30	1.1985757320361\\
-24	1.36107416703043\\
-18	1.42589848121563\\
-12	1.70144742966244\\
-6	1.92046816608733\\
0	2.01699612343231\\
6	1.82719505675379\\
12	1.47805537187922\\
18	1.34487628977399\\
24	1.14119448240936\\
};
\addlegendentry{1C}

\end{axis}
\end{tikzpicture}%
	} \\
	\caption{The estimated data rate $T$ versus the BS (TX) beam misalignment $\Delta_{TX}$ with respect to the best possible link (assuming optimal $\phi_{RX}$, $\theta_{RX}$). \label{fig:txmisalign}}
\end{figure}

A BS beam misalignment may occur due to a user moving out of the BS's beam (\emph{cf.}~Fig.~\ref{subfig:scenariotxrxmis}) or imperfect beam training (\emph{cf.}~Fig.~\ref{subfig:scenariotxrxmis}). In this section, we study the residual loss in estimated achievable data rate for such BS beam misalignments, assuming that the UE locally adjusts its orientation dynamically to mitigate the BS beam misalignment effects (\emph{cf.}~Figs.~\ref{subfig:scenariotxmisnomove} and~\ref{subfig:scenariotxmis}). Namely, we consider
\begin{eqnarray}
	T(\Delta_{TX}) = \underset{\phi_{RX}, \theta_{RX}}{\mathrm{max}} \; T( \Delta_{TX}, \phi_{RX}, \theta_{RX} ), 
\end{eqnarray}
where $\Delta_{TX}$ is the beam misalignment with respect to the best BS orientation $\phi_{TX,best}$, i.e.
\begin{eqnarray}
\Delta_{TX} &=& \phi_{TX}-\phi_{TX,best},\\
\label{eq:bestorientation}
\phi_{TX,best} &=&  \mathrm{arg\; max} \; T(\phi_{TX}, \phi_{RX}, \theta_{RX}).
\end{eqnarray}
We neglect the elevation misalignment assuming that all UEs are in the same plane, such that the BS's elevation beam misalignment is negligible as compared to its HPBW of $36^\circ$. 

In Fig.~\ref{fig:txmisalign} we show the estimated achievable data rate $T$ of a link over the BS (TX) beam misalignment $\Delta_{TX}$ with respect to the best BS orientation $\phi_{TX,best}$. Based on Figs.~\ref{subfig:txmisalign_trx_tx1}--\ref{subfig:txmisalign_trx_tx2} we make three observations for the phased antenna array results. First, we observe that the maximum data rate is only achieved without beam misalignment, i.e. there is only one maximum peak. Second, small beam misalignments in the order of the HPBW cause significant losses in data rate of up to 70\%, e.g. a beam misalignment of $-7^\circ$ in scenario \emph{1A} produces a drop from 4 Gbps to 1.4~Gbps and from 2.6~Gbps to 0.8~Gbps in scenario 2D. Third, the loss in data rate is not increasing monotonically with increasing beam misalignment. Overall, the results in Figs.~\ref{subfig:txmisalign_trx_tx1}--\ref{subfig:txmisalign_trx_tx2} demonstrate that a UE locally adjusting its orientation is not sufficient to fully mitigate the effects of BS beam misalignment. Consequently, strategies for tracking UE movement at the BS are required to maintain high data rates. Moreover, given the observed non-monotonic relationship between beam misalignment and loss in data rate, we expect that some types of beam tracking algorithms will not work reliably. For instance, let us consider an RSS-gradient-following beam tracking approach, i.e. an algorithm that tries to find the ideal beam orientation by adapting the beam orientation of its phased antenna array beam to a neighboring orientation that achieves a higher RSS. Due to the observed irregular variation of the achievable data rate over increasing beam misalignment, it would not be able to find the global optimal orientation.

In Fig.~\ref{subfig:txmisalign_horn}, we present the estimated maximum data rate over the horn antenna BS beam misalignment for comparison with the phased antenna array results. We observe that the data rates decrease monotonically with increasing misalignment which is in contrast to the observations made in Figs.~\mbox{\ref{subfig:txmisalign_trx_tx1}--\ref{subfig:txmisalign_trx_tx2}} for the phased antenna array. We emphasize that this stark dissimilarity between the results for the antenna types indicates that we cannot simply infer the impact of beam misalignments on mm-wave links based on measurements taken with horn antennas as in e.g. \cite{simic_60_2016, lee_field-measurement-based_2018} or calculations based on simplistic directional antenna radiation pattern models as in e.g.~\cite{wildman_joint_2014}. Instead, realistic phased antenna array radiation patterns have to be explicitly taken into account in the design of beam training and tracking algorithms.

\subsubsection{UE Beam Misalignment Analysis}
\label{subsec:rxmisalignment}

\begin{figure}
	\centering
	\subfloat[Measurement \textit{1A}.\label{subfig:rxmisalign_tx1rxa}]			{
		\import{tikz/LossInDatarate/}{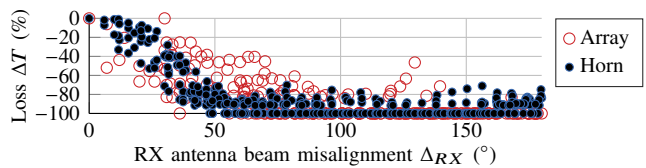}
	} \\
	\subfloat[Measurement \textit{1B}.\label{subfig:rxmisalign_tx1rxb}]			{
		\import{tikz/LossInDatarate/}{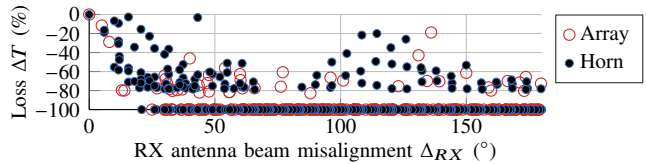}
	} \\
	\subfloat[Measurement \textit{1C}.\label{subfig:rxmisalign_tx1rxc}]			{
		\import{tikz/LossInDatarate/}{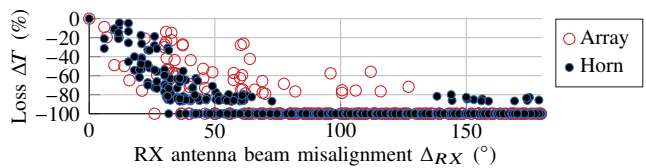}
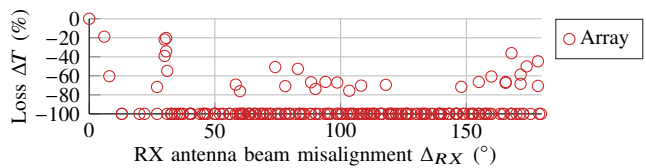
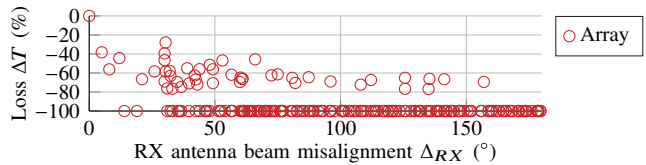
	} \\
	\subfloat[Measurement \textit{2B}.\label{subfig:rxmisalign_tx2rxb}]			{
%
%
\hspace*{-11pt}
\begin{tikzpicture}

\begin{axis}[%
width=171.166pt,
height=36pt,
at={(0pt,0pt)},
scale only axis,
xmin=0,
xmax=180,
xlabel near ticks,
xlabel shift={-4pt},
xlabel style={font={\footnotesize\color{black!100}}},
xlabel={$\text{RX antenna beam misalignment } \Delta_{RX} \text{ (}^\circ\text{)}$},
ymin=-100,
ymax=0,
ylabel near ticks,
ylabel shift={-5pt},
ylabel style={font={\footnotesize\color{black!100}}},
ylabel={Loss $\Delta T$ (\%)},
axis background/.style={fill=white},
axis x line*=bottom,
axis y line*=left,
xmajorgrids,
ymajorgrids,
legend style={legend cell align=left, align=left, draw=white!15!black},
ticklabel style={font={\color{black}\footnotesize}},legend style={font={\color{black}\footnotesize}},legend pos=outer north east,
]
\addplot[only marks, mark=o, mark options={}, mark size=2.1213pt, draw=red] table[row sep=crcr]{%
x	y\\
103.4456378974	-100\\
99	-100\\
103.4456378974	-75.6510493250495\\
115.762688289448	-100\\
98.6711710683521	-100\\
94	-66.3948352477308\\
98.6711710683521	-67.0022485918346\\
111.516814875605	-100\\
90.1387818865997	-100\\
85	-100\\
90.1387818865997	-100\\
104.043260233424	-100\\
83.5703296631047	-100\\
78	-100\\
83.5703296631047	-100\\
98.4073168011404	-100\\
78	-100\\
72	-100\\
78	-70.8299122414118\\
93.7229961108798	-100\\
70.6823881882892	-100\\
64	-100\\
70.6823881882892	-100\\
87.7268487978452	-100\\
66.1891229734916	-100\\
59	-100\\
66.1891229734916	-100\\
84.1486779456457	-100\\
60.0333240792145	-100\\
52	-100\\
60.0333240792145	-100\\
79.3977329651168	-100\\
50	-100\\
40	-100\\
50	-100\\
72.1110255092798	-100\\
46.0977222864644	-100\\
35	-100\\
46.0977222864644	-100\\
69.462219947249	-100\\
40.3608721412211	-100\\
27	-71.7025877454195\\
40.3608721412211	-100\\
65.7951365983839	-100\\
37.2021504754766	-100\\
22	-100\\
37.2021504754766	-100\\
63.9061812346818	-100\\
32.6955654485436	-100\\
13	-100\\
32.6955654485436	-100\\
61.3921819126833	-100\\
30.5941170815567	-34.2184089562754\\
6	-18.8142636808841\\
30.5941170815567	-20.2021188028466\\
60.2992537267253	-100\\
30	-39.0830291973797\\
0	0\\
30	-21.7978106022678\\
60	-76.1223853241588\\
31.04834939252	-100\\
8	-60.4179959682809\\
31.04834939252	-54.785285880411\\
60.5309838016862	-100\\
32.6955654485436	-100\\
13	-100\\
32.6955654485436	-100\\
61.3921819126833	-100\\
36.0555127546399	-100\\
20	-100\\
36.0555127546399	-100\\
63.2455532033676	-100\\
40.3608721412211	-100\\
27	-100\\
40.3608721412211	-100\\
65.7951365983839	-100\\
45.3431361950185	-100\\
34	-100\\
45.3431361950185	-100\\
68.9637585982667	-100\\
50	-100\\
40	-100\\
50	-100\\
72.1110255092798	-100\\
53.2541078227774	-100\\
44	-100\\
53.2541078227774	-100\\
74.4043009509531	-100\\
55.7584074378026	-100\\
47	-100\\
55.7584074378026	-100\\
76.2167960491649	-100\\
58.309518948453	-100\\
50	-100\\
58.309518948453	-69.2079735288846\\
78.1024967590665	-100\\
63.5295206970744	-100\\
56	-100\\
63.5295206970744	-100\\
82.0731381147328	-100\\
67.9779375974294	-100\\
61	-100\\
67.9779375974294	-100\\
85.5628424025289	-100\\
75.2396172239067	-100\\
69	-100\\
75.2396172239067	-100\\
91.438503924769	-100\\
79.8498591107085	-100\\
74	-50.7861149916568\\
79.8498591107085	-100\\
95.2680429105164	-100\\
88.2553114548921	-100\\
83	-52.6289666864277\\
88.2553114548921	-66.7403664706018\\
102.415819090607	-100\\
94.8683298050514	-100\\
90	-73.7939469614621\\
94.8683298050514	-100\\
108.16653826392	-70.2791682030774\\
100.578327685441	-100\\
96	-100\\
100.578327685441	-100\\
113.207773584679	-100\\
108.240473021878	-100\\
104	-100\\
108.240473021878	-100\\
120.066648158429	-100\\
113.053084876088	-100\\
109	-100\\
113.053084876088	-100\\
124.422666745252	-100\\
119.816526406001	-100\\
116	-100\\
119.816526406001	-100\\
130.598621738516	-100\\
126.605687076055	-100\\
123	-100\\
126.605687076055	-100\\
136.853936735485	-100\\
133.416640641263	-100\\
130	-100\\
133.416640641263	-100\\
143.178210632764	-100\\
139.269522868429	-100\\
136	-100\\
139.269522868429	-100\\
148.647233408496	-100\\
143.178210632764	-100\\
140	-100\\
143.178210632764	-100\\
152.315462117278	-100\\
146.1129699924	-100\\
143	-100\\
146.1129699924	-100\\
155.077400029792	-100\\
149.050327071094	-100\\
146	-100\\
149.050327071094	-100\\
157.848028179005	-100\\
154.932243254915	-100\\
152	-100\\
154.932243254915	-100\\
163.413585726524	-100\\
159.840545544615	-100\\
157	-100\\
159.840545544615	-100\\
168.074388292803	-100\\
167.705098312484	-100\\
165	-100\\
167.705098312484	-100\\
175.570498660794	-100\\
172.626765016321	-100\\
170	-100\\
172.626765016321	-100\\
179.722436226801	-100\\
178.503443558838	-70.5781682704489\\
179	-100\\
178.503443558838	-44.6869110390191\\
171.211758840758	-100\\
171.596178382709	-58.7053652022435\\
174	-50.069308975876\\
171.596178382709	-68.3801556786367\\
164.562030301172	-100\\
165.67038311158	-67.3312355358386\\
168	-36.143351725115\\
165.67038311158	-66.4739149834653\\
158.843344629117	-100\\
157.762515838433	-100\\
160	-60.7197685379227\\
157.762515838433	-100\\
151.193869821789	-100\\
152.816506448993	-100\\
155	-66.482051476519\\
152.816506448993	-100\\
146.399906367062	-100\\
145.887879838623	-100\\
148	-71.5918685626373\\
145.887879838623	-100\\
139.672970337273	-100\\
138.954755762536	-100\\
141	-100\\
138.954755762536	-100\\
132.929526358005	-100\\
132.017544534673	-100\\
134	-100\\
132.017544534673	-100\\
126.171002653649	-100\\
126.068386061225	-100\\
128	-100\\
126.068386061225	-100\\
120.366947187997	-100\\
122.10086170816	-100\\
124	-100\\
122.10086170816	-100\\
116.492299916409	-100\\
119.124513492967	-100\\
121	-100\\
119.124513492967	-100\\
113.583685605031	-100\\
116.147585617858	-100\\
118	-69.4479754162051\\
116.147585617858	-100\\
110.672905603904	-100\\
110.192073784678	-100\\
112	-100\\
110.192073784678	-100\\
104.845145058927	-100\\
105.227552510088	-100\\
107	-100\\
105.227552510088	-100\\
99.9826928837236	-100\\
};
\addlegendentry{Array}

\end{axis}
\end{tikzpicture}%
	} \\
	\subfloat[Measurement \textit{2D}.\label{subfig:rxmisalign_tx2rxd}]			{
%
%
\hspace*{-11pt}
\begin{tikzpicture}

\begin{axis}[%
width=171.166pt,
height=36pt,
at={(0pt,0pt)},
scale only axis,
xmin=0,
xmax=180,
xlabel near ticks,
xlabel shift={-4pt},
xlabel style={font={\footnotesize\color{black!100}}},
xlabel={$\text{RX antenna beam misalignment } \Delta_{RX} \text{ (}^\circ\text{)}$},
ymin=-100,
ymax=0,
ylabel near ticks,
ylabel shift={-5pt},
ylabel style={font={\footnotesize\color{black!100}}},
ylabel={Loss $\Delta T$ (\%)},
axis background/.style={fill=white},
axis x line*=bottom,
axis y line*=left,
xmajorgrids,
ymajorgrids,
legend style={legend cell align=left, align=left, draw=white!15!black},
ticklabel style={font={\color{black}\footnotesize}},legend style={font={\color{black}\footnotesize}},legend pos=outer north east,
]
\addplot[only marks, mark=o, mark options={}, mark size=2.1213pt, draw=red] table[row sep=crcr]{%
x	y\\
178.538511251774	-100\\
176	-100\\
178.538511251774	-100\\
174.053771213289	-100\\
170.657551839935	-100\\
168	-100\\
170.657551839935	-100\\
178.392824967822	-100\\
165.737744644966	-100\\
163	-100\\
165.737744644966	-100\\
173.692256591939	-100\\
158.858427538485	-100\\
156	-100\\
158.858427538485	-100\\
167.140659326209	-100\\
147.091808065575	-100\\
144	-100\\
147.091808065575	-100\\
156	-100\\
142.200562586791	-100\\
139	-100\\
142.200562586791	-100\\
151.396829557293	-100\\
134.391219951305	-100\\
131	-100\\
134.391219951305	-100\\
144.086779407411	-100\\
129.522198869537	-100\\
126	-100\\
129.522198869537	-100\\
139.556440195356	-100\\
120.784932835184	-100\\
117	-100\\
120.784932835184	-100\\
131.487642004867	-100\\
114.017542509914	-100\\
110	-100\\
114.017542509914	-100\\
125.299640861417	-100\\
108.240473021878	-100\\
104	-100\\
108.240473021878	-100\\
120.066648158429	-100\\
100.578327685441	-100\\
96	-100\\
100.578327685441	-100\\
113.207773584679	-100\\
95.8175349296777	-100\\
91	-100\\
95.8175349296777	-100\\
109	-100\\
89.196412483911	-100\\
84	-100\\
89.196412483911	-100\\
103.227903204512	-100\\
82.6377637645163	-100\\
77	-100\\
82.6377637645163	-100\\
97.6165969494942	-100\\
76.1577310586391	-100\\
70	-100\\
76.1577310586391	-100\\
92.1954445729289	-100\\
70.6823881882892	-100\\
64	-100\\
70.6823881882892	-100\\
87.7268487978452	-100\\
67.0820393249937	-100\\
60	-100\\
67.0820393249937	-100\\
84.8528137423857	-100\\
64.4127316607517	-100\\
57	-100\\
64.4127316607517	-100\\
82.7586853448023	-100\\
61.773780845922	-100\\
54	-100\\
61.773780845922	-100\\
80.7217442824423	-100\\
56.6038867923396	-100\\
48	-100\\
56.6038867923396	-100\\
76.8374908491942	-100\\
52.4309069156733	-100\\
43	-72.1103299861221\\
52.4309069156733	-100\\
73.8173421358423	-100\\
46.0977222864644	-100\\
35	-69.4496571682587\\
46.0977222864644	-100\\
69.462219947249	-100\\
42.4264068711929	-100\\
30	-39.404286673151\\
42.4264068711929	-66.7613229850283\\
67.0820393249937	-100\\
36.6196668472011	-100\\
21	-66.5758885674735\\
36.6196668472011	-74.5904120209249\\
63.5688603012513	-100\\
33.1058907144937	-100\\
14	-100\\
33.1058907144937	-76.5608042069653\\
61.6116872029975	-100\\
31.04834939252	-100\\
8	-56.1861094162349\\
31.04834939252	-76.114657360112\\
60.5309838016862	-100\\
30	-69.016804832777\\
0	0\\
30	-46.4939121501089\\
60	-69.5269898176977\\
30.4138126514911	-58.3654040114952\\
5	-38.2761852908889\\
30.4138126514911	-27.891992779778\\
60.2079728939615	-64.8463368640845\\
32.310988842807	-63.0564350940944\\
12	-44.3331834006986\\
32.310988842807	-100\\
61.1882341631134	-100\\
35.5105618091294	-100\\
19	-100\\
35.5105618091294	-100\\
62.9364759102383	-100\\
39.6988664825584	-70.9568172945933\\
26	-58.3454075148583\\
39.6988664825584	-100\\
65.3911308970873	-100\\
43.8634243989226	-56.0011209329181\\
32	-57.8462015943936\\
43.8634243989226	-100\\
68	-100\\
46.8614980554399	-100\\
36	-100\\
46.8614980554399	-100\\
69.9714227381436	-100\\
49.2036584005702	-56.1249929395727\\
39	-55.0413521855083\\
49.2036584005702	-70.7057860283063\\
71.5611626512594	-100\\
51.6139516022558	-100\\
42	-62.5448504618079\\
51.6139516022558	-100\\
73.2393336944022	-100\\
56.6038867923396	-61.8707024125935\\
48	-51.4085488798334\\
56.6038867923396	-100\\
76.8374908491942	-100\\
60.9015599143405	-66.2702519147854\\
53	-46.8327779716976\\
60.9015599143405	-100\\
80.0562302385017	-100\\
67.9779375974294	-100\\
61	-66.0096659529445\\
67.9779375974294	-100\\
85.5628424025289	-100\\
72.4982758415674	-62.491965910104\\
66	-45.6188690509748\\
72.4982758415674	-100\\
89.196412483911	-100\\
80.7774721070176	-65.1004028787147\\
75	-61.4224389635287\\
80.7774721070176	-100\\
96.0468635614927	-100\\
87.3155198117723	-64.4863856374504\\
82	-70.4753094522723\\
87.3155198117723	-100\\
101.607086367044	-100\\
92.9731143933557	-100\\
88	-100\\
92.9731143933557	-100\\
106.508215645555	-100\\
100.578327685441	-100\\
96	-68.9788819841513\\
100.578327685441	-100\\
113.207773584679	-100\\
105.361283211624	-100\\
101	-100\\
105.361283211624	-100\\
117.477657450257	-100\\
112.089250153616	-67.3946548625334\\
108	-72.3144817684521\\
112.089250153616	-100\\
123.547561691844	-100\\
118.848643240047	-100\\
115	-100\\
118.848643240047	-100\\
129.711217710728	-100\\
125.634390196315	-65.12526968839\\
122	-100\\
125.634390196315	-76.6513750402932\\
135.955875194859	-100\\
131.468627436358	-100\\
128	-100\\
131.468627436358	-100\\
141.364776376578	-100\\
135.366170072142	-66.1673227607067\\
132	-100\\
135.366170072142	-100\\
144.996551683135	-100\\
138.293166859393	-100\\
135	-76.8221168957699\\
138.293166859393	-100\\
147.732867026942	-100\\
141.223227551278	-66.5313494717875\\
138	-100\\
141.223227551278	-100\\
150.479234447813	-100\\
147.091808065575	-100\\
144	-100\\
147.091808065575	-100\\
156	-100\\
151.990131258579	-100\\
149	-100\\
151.990131258579	-100\\
160.626896875959	-100\\
159.840545544615	-100\\
157	-69.3555226038665\\
159.840545544615	-100\\
168.074388292803	-100\\
164.754362612952	-100\\
162	-100\\
164.754362612952	-100\\
172.754160586656	-100\\
173.611635554763	-100\\
171	-100\\
173.611635554763	-100\\
178.779140273533	-100\\
179.489612487259	-100\\
178	-100\\
179.489612487259	-100\\
172.159642249063	-100\\
};
\addlegendentry{Array}

\end{axis}
\end{tikzpicture}%
	} 
	\caption{Loss in estimated data rate $T$ versus UE (RX) beam misalignment $\Delta_{RX}$ with respect to the best possible link (assuming optimal $\phi_{TX}$). \label{fig:rxmisalign}}
\end{figure}

In Figs.~\ref{subfig:scenariotxmisnomove}--\ref{subfig:scenariotxrxmis} we illustrated how imperfect beamtraining, a rotation, or a small-scale movement of the UE may lead to a UE beam misalignment. In this section, we study the effects of such UE beam misalignments, assuming that the BS locally adjusts its orientation dynamically to mitigate the UE beam misalignment effects. We thus apply the same principle of considering one-sided misalignment only, as in the previous analysis of the BS beam misalignment. Namely, we consider
\begin{eqnarray}
T(\Delta_{RX}) = \underset{\phi_{TX}}{\mathrm{max}} \; T( \phi_{TX}, \Delta_{RX} ).
\end{eqnarray}
where $\Delta_{RX}$ is the misalignment with respect to the best UE orientation $(\phi_{RX,best}, \theta_{RX,best})$, i.e.
\begin{eqnarray}
\Delta_{RX} = \sqrt{(\phi_{RX}-\phi_{RX,best})^2 + (\theta_{RX}-\theta_{RX,best})^2},\\
(\phi_{RX,best}, \theta_{RX, best}) =  \mathrm{arg\; max} \; T(\phi_{TX}, \phi_{RX}, \theta_{RX}).
\end{eqnarray}

In Fig.~\ref{fig:rxmisalign} we present the loss in estimated data rate $\Delta T$ for a UE (RX) antenna beam misalignment $\Delta_{RX}$ with respect to the best UE orientation. Figs.~\mbox{\ref{subfig:rxmisalign_tx1rxa}--\ref{subfig:rxmisalign_tx1rxc}} show results for both measurement setups, where we again observe that a misalignment causes a gradual loss in data rate for the horn antenna results, whereas the phased antenna array results vary rather irregularly. This confirms that we cannot simply infer the impact of beam misalignments on mm-wave links based on measurements taken with horn antennas. Moreover, in Figs.~\ref{subfig:rxmisalign_tx1rxa}--\ref{subfig:rxmisalign_tx2rxd} we find that a small misalignment of the UE phased antenna array beam of up to 8$^\circ$ results in data rate losses of 6--60\%. This demonstrates that even with perfect UE tracking, agile beamsteering at the UE is needed to compensate residual losses in data rate due to misalignment.


One key to enabling mm-wave networks is to successfully integrate UE phased antenna arrays that can reliably maintain a connection over all 3D orientations. As described in Sec.~\ref{sec:PhasedAntennaSetup}, the phased antenna array used in our measurements only allows azimuth beamsteering while having a wide elevation HPBW of $36^\circ$. To investigate if this is a suitable configuration for a mm-wave UE, we isolate the effect of the UE elevation angle on estimated data rates. We expect that the best data rate would be achieved for the measured elevation angles $\theta_{RX}$ that are closest to the ideal elevation orientation towards the BS, which can be computed from the distances between BS and UE (\emph{cf.}~Table~\ref{tab:measurements}). Namely, we expect the scenarios \emph{1A} and \emph{2D} to achieve approximately equal maximum estimated achievable data rate for $\theta_{RX}=0^\circ$ and $\theta_{RX}=30^\circ$, as the distances between BS and UE were the shortest in these cases, implying a greater elevation angle. For larger distances between BS and UE, the ideal elevation angle is smaller and we expect to see the maximum estimated achievable data rate for the scenarios \emph{1B}, \emph{1C}, and \emph{2B} at~$\theta_{RX}=0^\circ$.

\begin{figure}
	\centering
	\import{tikz/RXElev/}{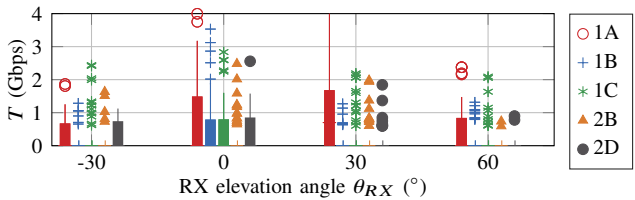}
	\caption{Boxplots of the estimated data rate $T$ that can be achieved with the phased array over all UE azimuth angles $\phi_{RX}$ with a fixed UE elevation angle $\theta_{RX}$ (assuming optimal $\phi_{TX}$).\label{fig:rxelev}}
\end{figure}

Fig.~\ref{fig:rxelev} shows boxplots of the estimated data rate $T$ over all azimuth angles \mbox{$\phi_{RX} \in [-180^\circ, 180^\circ]$} for UE elevation angle $\theta_{RX}$, i.e. we consider
\begin{eqnarray}
T(\theta_{RX}) = \underset{\phi_{TX}}{\mathrm{max}} \; T(\phi_{TX}, \phi_{RX}, \theta_{RX}). 
\end{eqnarray}
We observe a drop in maximum estimated data rate for all results except \emph{1A} when the elevation angle is not equal to 0$^\circ$. This is expected, except for the result \emph{2D} which shows an outlier achieving around 2.6~Gbps for $\theta_{RX}=0^\circ$, possibly due to the tree foliage that blocked the LOS (\emph{cf.}~Fig.~\ref{subfig:txrxposphotos_tx2rxd}). Overall, despite the large elevation HPBW of the phased antenna array pattern of $36^\circ$, the loss in the maximum achievable estimated data rate ranges between 26\% and 70\% for a change in elevation angle $\theta_{RX}$ from 0$^\circ$ to 60$^\circ$ (\emph{cf.} Fig.~\ref{fig:rxelev}). We conclude that 2D beamsteering will not be sufficient to fully exploit available UE link opportunities in a 3D environment. 


\subsubsection{Two-sided Beam Misalignment Analysis}
\label{subsubsec:rxbeamerrorfixedtx}

\begin{figure}[t]
	\centering
	
	\subfloat[Measurement \emph{1A}.\label{subfig:rxssmisalignment_tx1rxa}]{
%
%
\hspace*{-11pt}
\begin{tikzpicture}

\begin{axis}[%
width=170pt,
height=45pt,
at={(0pt,0pt)},
scale only axis,
xmin=-180,
xmax=180,
xlabel near ticks,
xlabel shift={-4pt},
xlabel style={font={\footnotesize\color{black!100}}},
xlabel={$\text{RX azimuth angle } \phi_{RX}\text{ (}^\circ\text{)}$},
ymin=0,
ymax=4,
ylabel near ticks,
ylabel shift={-5pt},
ylabel style={font={\footnotesize\color{black!100}}},
ylabel={$T$ (Gbps)},
axis background/.style={fill=white},
xmajorgrids,
ymajorgrids,
legend style={at={(1.03,0.5)}, anchor=west, legend cell align=left, align=left, draw=white!15!black},
ticklabel style={font={\color{black}\footnotesize}},legend style={font={\color{black}\footnotesize}},legend pos=outer north east,
]
\addlegendimage{color=white}
\addlegendentry{$\phi{}_{\text{TX}}$};
\addplot [color=red, densely dashed, mark=o, mark options={solid, red}]
  table[row sep=crcr]{%
-175	0\\
-168	0\\
-156	0\\
-151	0\\
-129	0\\
-116	0\\
-103	0\\
-96	0\\
-89	0\\
-82	0\\
-76	0\\
-72	0\\
-69	0\\
-66	0\\
-60	0\\
-55	0\\
-47	0\\
-42	0\\
-33	0\\
-26	0\\
-20	0\\
-12	0\\
-7	1.36370823351449\\
0	1.35317615738779\\
7	0\\
14	0\\
20	0\\
24	0\\
27	0\\
30	0\\
36	0\\
41	0\\
49	0\\
54	0\\
63	0\\
70	0\\
76	0\\
84	0\\
89	0\\
96	0\\
103	0\\
110	0\\
116	0\\
120	0\\
123	0\\
126	0\\
132	0\\
137	0\\
145	0\\
150	0\\
159	0\\
166	0\\
172	0\\
180	0\\
};
\addlegendentry{$-7^\circ$}

\addplot [color=blue, mark=triangle*, mark options={solid, blue}]
  table[row sep=crcr]{%
-175	0\\
-168	0\\
-156	0\\
-151	0\\
-143	0\\
-138	0\\
-129	0\\
-122	0\\
-116	0\\
-108	0\\
-103	0\\
-96	0\\
-89	0\\
-82	0\\
-76	1.00469068223461\\
-72	0\\
-69	1.00926694864327\\
-66	0\\
-60	1.0192192866608\\
-55	1.83694668283956\\
-47	2.13120847906816\\
-42	2.02033665200862\\
-33	2.44951122594182\\
-26	1.18439051116425\\
-20	0\\
-12	2.24720665711687\\
-7	3.75211116387939\\
0	3.99492688965108\\
7	1.86225155836556\\
14	3.17543731607048\\
20	1.99756013310831\\
24	1.78389540508526\\
27	2.00407552807685\\
30	2.64830950827746\\
36	2.01675040977717\\
41	1.49063067239021\\
49	0.947205498356665\\
54	0\\
63	1.03699113534948\\
70	0\\
76	0\\
84	0\\
89	0\\
96	0\\
103	0\\
110	0.826335225711852\\
116	0\\
120	0.95127676648238\\
123	0\\
126	1.46258551558633\\
132	0\\
137	0\\
145	0\\
150	0\\
159	0\\
166	0\\
172	0\\
180	0\\
};
\addlegendentry{$0^\circ$}

\addplot [color=green, solid, mark=asterisk, mark options={solid, green}]
  table[row sep=crcr]{%
-175	0\\
-168	0.674483320027176\\
-156	0\\
-151	0\\
-143	0\\
-138	0\\
-129	0\\
-122	0\\
-116	0\\
-108	0\\
-103	0\\
-96	0\\
-89	0\\
-82	0\\
-76	0\\
-72	0\\
-69	0\\
-66	0\\
-60	0\\
-55	0.921800412373855\\
-47	1.27261565983868\\
-42	1.35174591237031\\
-33	1.52434425694919\\
-26	0\\
-20	0\\
-12	1.25631011980894\\
-7	2.31986264357755\\
0	3.33357525867297\\
7	0\\
14	2.23343696435048\\
20	1.1934935039033\\
24	0.832841201926091\\
27	1.18462177119316\\
30	1.30612385770695\\
36	1.17180643672011\\
41	1.07806734083872\\
49	0\\
54	0.629669447830449\\
63	0\\
70	0\\
76	0\\
84	0\\
89	0\\
96	0\\
103	0\\
110	0\\
116	0\\
120	0\\
123	0\\
126	0.760300949187331\\
132	0\\
137	0\\
145	0\\
150	0\\
159	0\\
166	0\\
172	0\\
180	0\\
};
\addlegendentry{$7^\circ$}

\end{axis}

\begin{axis}[%
width=224.577pt,
height=49.906pt,
at={(-23.269pt,-6.163pt)},
scale only axis,
xmin=0,
xmax=1,
ymin=0,
ymax=1,
axis line style={draw=none},
ticks=none,
axis x line*=bottom,
axis y line*=left,
legend style={legend cell align=left, align=left, draw=white!15!black},
ticklabel style={font={\color{black}\footnotesize}},legend style={font={\color{black}\footnotesize}},legend pos=outer north east,
]
\end{axis}
\end{tikzpicture}%
	} \\
	\subfloat[Measurement \emph{1B}.\label{subfig:rxssmisalignment_tx1rxb}]{
%
%
\begin{tikzpicture}

\begin{axis}[%
width=170pt,
height=45pt,
at={(0pt,0pt)},
scale only axis,
xmin=-180,
xmax=180,
xlabel near ticks,
xlabel shift={-4pt},
xlabel style={font={\footnotesize\color{black!100}}},
xlabel={$\text{RX azimuth angle } \phi_{RX}\text{ (}^\circ\text{)}$},
ymin=0,
ymax=4,
ylabel near ticks,
ylabel shift={-5pt},
ylabel style={font={\footnotesize\color{black!100}}},
ylabel={$T$ (Gbps)},
axis background/.style={fill=white},
xmajorgrids,
ymajorgrids,
legend style={at={(1.03,0.5)}, anchor=west, legend cell align=left, align=left, draw=white!15!black},
ticklabel style={font={\color{black}\footnotesize}},legend style={font={\color{black}\footnotesize}},legend pos=outer north east,
]
\addlegendimage{color=white}
\addlegendentry{$\phi{}_{\text{TX}}$};
\addplot [color=red, densely dashed, mark=o, mark options={solid, red}]
  table[row sep=crcr]{%
-173	0\\
-168	0\\
-161	0\\
-149	0\\
-144	0.808540192303482\\
-136	1.29885586709127\\
-131	1.1111170286324\\
-122	0\\
-115	0\\
-109	0\\
-101	0\\
-96	0\\
-89	0\\
-82	0\\
-75	0\\
-69	0\\
-65	0\\
-62	0\\
-59	0\\
-53	0\\
-48	0\\
-40	0\\
-35	0\\
-26	0\\
-13	0\\
-5	0.80524323047303\\
0	1.41931601648163\\
7	0\\
14	0.681494629490286\\
21	0\\
27	0\\
31	0\\
34	0\\
37	0\\
43	0\\
48	0\\
56	0\\
61	0\\
70	0\\
77	0\\
103	0\\
110	0\\
117	0\\
123	0\\
130	0\\
133	0\\
139	0\\
144	0\\
152	0\\
157	0\\
166	0\\
173	0\\
179	0\\
};
\addlegendentry{$-7^\circ$}

\addplot [color=blue, mark=triangle*, mark options={solid, blue}]
  table[row sep=crcr]{%
-173	0.749124650859707\\
-168	0.76940465202258\\
-161	0\\
-149	0.714447740381801\\
-144	2.51157937320422\\
-136	3.53106559421553\\
-131	3.12234377689941\\
-122	0.709581006305474\\
-115	1.01195172981178\\
-109	0.791959322153573\\
-101	0.722241379676748\\
-96	1.90050492248599\\
-89	0.982963675247237\\
-82	1.54321677245093\\
-75	0.788718260756615\\
-69	0\\
-65	0\\
-62	0\\
-59	1.38724612063878\\
-53	0\\
-48	0.606788941966817\\
-40	1.18642269264746\\
-35	0.751111296838567\\
-26	0\\
-19	0\\
-13	0.867663257029826\\
-5	2.01605860837208\\
0	2.86371016869504\\
7	0\\
14	1.36451152786012\\
21	0\\
27	0\\
31	0\\
34	0\\
37	0.696697341902916\\
43	0\\
48	0\\
56	0\\
61	0\\
70	0\\
77	0\\
83	0\\
91	0\\
96	0\\
103	0\\
110	0\\
117	0\\
123	0\\
127	0\\
130	0\\
133	0\\
139	0\\
144	0\\
152	0\\
157	0\\
166	0\\
173	0\\
179	0\\
};
\addlegendentry{$0^\circ$}

\addplot [color=green, mark=asterisk, mark options={solid, green}]
  table[row sep=crcr]{%
-173	0\\
-168	0\\
-161	0\\
-149	0\\
-144	0.807711497313471\\
-136	1.61646444098858\\
-131	1.23970557219452\\
-122	0\\
-115	0\\
-109	0\\
-101	0\\
-96	0\\
-89	0\\
-82	0\\
-75	0\\
-69	0\\
-65	0\\
-62	0\\
-59	0\\
-53	0\\
-48	0\\
-40	0\\
-35	0\\
-26	0\\
-19	0\\
-13	0\\
-5	0\\
0	0\\
7	0\\
14	0\\
21	0\\
27	0\\
31	0\\
34	0\\
37	0\\
43	0\\
48	0\\
56	0\\
61	0\\
70	0\\
77	0\\
83	0\\
91	0\\
96	0\\
103	0\\
110	0\\
117	0\\
123	0\\
127	0\\
130	0\\
133	0\\
139	0\\
144	0\\
157	0\\
166	0\\
173	0\\
179	0\\
};
\addlegendentry{$7^\circ$}

\end{axis}
\end{tikzpicture}%
	} \\
	\caption{Estimated data rate $T$ versus RX azimuth angles $\phi_{RX}$ for selected measurements and selected fixed BS orientations $\phi_{TX}$, representative of small-scale BS beam misalignment ($\theta_{RX}=0^\circ$ assumed throughout).\label{fig:rxssmisalignment}}
\end{figure}
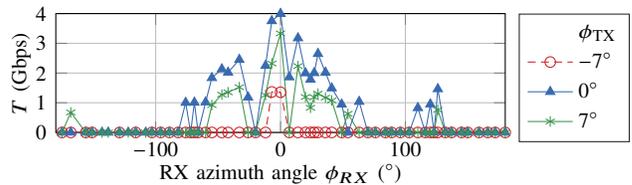
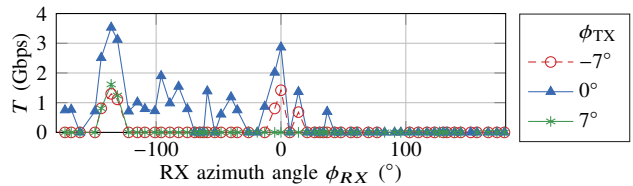

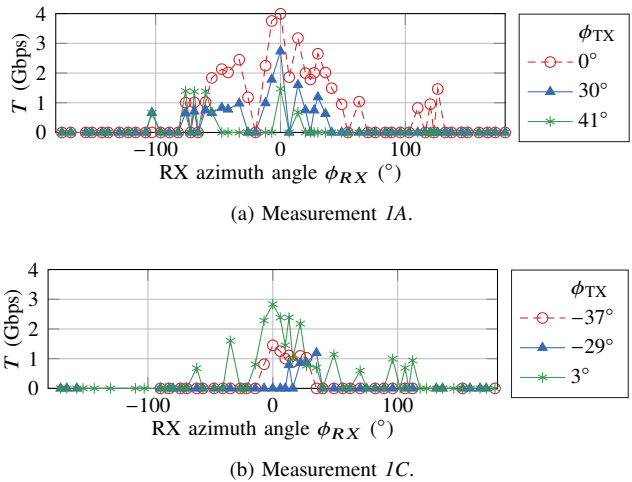
\begin{figure}[t]
	\centering
	\subfloat[Measurement \emph{1A}.\label{subfig:fixedtx_linkop_1a}]{
%
%
\hspace*{-11pt}
\begin{tikzpicture}

\begin{axis}[%
width=170pt,
height=45pt,
at={(0pt,0pt)},
scale only axis,
xmin=-180,
xmax=180,
xlabel near ticks,
xlabel shift={-4pt},
xlabel style={font={\footnotesize\color{black!100}}},
xlabel={$\text{RX azimuth angle } \phi_{RX}\text{ (}^\circ\text{)}$},
ymin=0,
ymax=4,
ylabel near ticks,
ylabel shift={-5pt},
ylabel style={font={\footnotesize\color{black!100}}},
ylabel={$T$ (Gbps)},
axis background/.style={fill=white},
xmajorgrids,
ymajorgrids,
legend style={legend cell align=left, align=left, draw=white!15!black},
ticklabel style={font={\color{black}\footnotesize}},legend style={font={\color{black}\footnotesize}},legend pos=outer north east,
]
\addlegendimage{color=white}
\addlegendentry{$\phi{}_{\text{TX}}$};
\addplot [color=red, densely dashed, mark=o, mark options={solid, red}]
  table[row sep=crcr]{%
-175	0\\
-168	0\\
-156	0\\
-151	0\\
-143	0\\
-138	0\\
-129	0\\
-122	0\\
-116	0\\
-108	0\\
-103	0\\
-96	0\\
-89	0\\
-82	0\\
-76	1.00469068223461\\
-72	0\\
-69	1.00926694864327\\
-66	0\\
-60	1.0192192866608\\
-55	1.83694668283956\\
-47	2.13120847906816\\
-42	2.02033665200862\\
-33	2.44951122594182\\
-26	1.18439051116425\\
-20	0\\
-12	2.24720665711687\\
-7	3.75211116387939\\
0	3.99492688965108\\
7	1.86225155836556\\
14	3.17543731607048\\
20	1.99756013310831\\
24	1.78389540508526\\
27	2.00407552807685\\
30	2.64830950827746\\
36	2.01675040977717\\
41	1.49063067239021\\
49	0.947205498356665\\
54	0\\
63	1.03699113534948\\
70	0\\
76	0\\
84	0\\
89	0\\
96	0\\
103	0\\
110	0.826335225711852\\
116	0\\
120	0.95127676648238\\
123	0\\
126	1.46258551558633\\
132	0\\
137	0\\
145	0\\
150	0\\
159	0\\
166	0\\
172	0\\
180	0\\
};
\addlegendentry{$0^\circ$}

\addplot [color=blue, mark=triangle*, mark options={solid, blue}]
  table[row sep=crcr]{%
-175	0\\
-168	0\\
-156	0\\
-151	0\\
-143	0\\
-138	0\\
-129	0\\
-122	0\\
-116	0\\
-108	0\\
-103	0.662029818913425\\
-96	0\\
-89	0\\
-82	0\\
-76	0.64926648515862\\
-72	0\\
-69	0.709528261814412\\
-66	0\\
-60	0.748755714327923\\
-55	0.660108340205154\\
-47	0.833954190903251\\
-42	0.776764753277292\\
-33	0.979429883922071\\
-26	0\\
-20	0\\
-12	0.990576128996886\\
-7	1.78344424013364\\
0	2.73347189459133\\
7	0\\
14	1.60197656815907\\
20	0.751236564490785\\
24	0\\
27	0.746336252710165\\
30	1.18790930023332\\
36	0.628036118911411\\
41	0\\
49	0\\
54	0\\
63	0\\
70	0\\
76	0\\
84	0\\
89	0\\
96	0\\
103	0\\
110	0\\
116	0\\
120	0\\
123	0\\
126	0\\
132	0\\
137	0\\
145	0\\
150	0\\
159	0\\
166	0\\
172	0\\
180	0\\
};
\addlegendentry{$30^\circ$}

\addplot [color=green, mark=asterisk, mark options={solid, green}]
  table[row sep=crcr]{%
-175	0\\
-168	0\\
-156	0\\
-151	0\\
-143	0\\
-138	0\\
-129	0\\
-122	0\\
-116	0\\
-108	0\\
-103	0.661478007915153\\
-96	0\\
-89	0\\
-82	0\\
-76	1.39499203381704\\
-72	0\\
-69	1.37816687284163\\
-66	0\\
-60	1.38720240098436\\
-55	0.6850098797811\\
-47	0\\
-42	0\\
-33	0\\
-26	0\\
-20	0\\
-12	0\\
-7	0\\
0	1.46882390218231\\
7	0\\
14	0.680742493056457\\
20	0\\
24	0\\
27	0\\
30	0\\
36	0\\
41	0\\
49	0\\
54	0\\
63	0\\
70	0\\
76	0\\
84	0\\
89	0\\
96	0\\
103	0\\
110	0\\
116	0\\
120	0\\
123	0\\
126	0\\
132	0\\
137	0\\
145	0\\
150	0\\
159	0\\
166	0\\
172	0\\
180	0\\
};
\addlegendentry{$41^\circ$}

\end{axis}

\begin{axis}[%
width=224.577pt,
height=49.906pt,
at={(-23.168pt,-6.163pt)},
scale only axis,
xmin=0,
xmax=1,
ymin=0,
ymax=1,
axis line style={draw=none},
ticks=none,
axis x line*=bottom,
axis y line*=left,
legend style={legend cell align=left, align=left, draw=white!15!black},
ticklabel style={font={\color{black}\footnotesize}},legend style={font={\color{black}\footnotesize}},legend pos=outer north east,
]
\end{axis}
\end{tikzpicture}%
	} \\
	\subfloat[Measurement \emph{1C}.\label{subfig:fixedtx_linkop_1c}]{
%
%
\hspace*{-11pt}
\begin{tikzpicture}

\begin{axis}[%
width=170pt,
height=45pt,
at={(0pt,0pt)},
scale only axis,
xmin=-180,
xmax=180,
xlabel near ticks,
xlabel shift={-4pt},
xlabel style={font={\footnotesize\color{black!100}}},
xlabel={$\text{RX azimuth angle } \phi_{RX}\text{ (}^\circ\text{)}$},
ymin=0,
ymax=4,
ylabel near ticks,
ylabel shift={-5pt},
ylabel style={font={\footnotesize\color{black!100}}},
ylabel={$T$ (Gbps)},
axis background/.style={fill=white},
xmajorgrids,
ymajorgrids,
legend style={legend cell align=left, align=left, draw=white!15!black},
ticklabel style={font={\color{black}\footnotesize}},legend style={font={\color{black}\footnotesize}},legend pos=outer north east,
]
\addlegendimage{color=white}
\addlegendentry{$\phi{}_{\text{TX}}$};
\addplot [color=red, densely dashed, mark=o, mark options={solid, red}]
  table[row sep=crcr]{%
-90	0\\
-83	0\\
-74	0\\
-69	0\\
-61	0\\
-56	0\\
-47	0\\
-40	0\\
-34	0\\
-26	0\\
-21	0\\
-14	0\\
-7	0.812827495966168\\
0	1.44959058110542\\
6	1.25289456379\\
10	1.00333153093403\\
13	1.10889736702195\\
16	0.890037971661608\\
22	1.09507743397488\\
27	1.03236634095271\\
35	0\\
40	0\\
49	0\\
56	0\\
62	0\\
70	0\\
82	0\\
89	0\\
96	0\\
106	0\\
112	0\\
152	0\\
178	0\\
};
\addlegendentry{$-37^\circ$}

\addplot [color=blue, mark=triangle*, mark options={solid, blue}]
  table[row sep=crcr]{%
-170	0\\
-165	0\\
-157	0\\
-90	0\\
-83	0\\
-80	0\\
-74	0\\
-69	0\\
-61	0\\
-56	0\\
-47	0\\
-40	0\\
-34	0\\
-26	0\\
-21	0\\
-14	0\\
-7	0\\
0	0\\
6	0\\
10	0\\
13	0.789128405854192\\
16	0\\
22	0.85612005462249\\
27	0.865549049710653\\
35	1.19380636986401\\
40	0\\
49	0\\
56	0\\
62	0\\
70	0\\
75	0\\
82	0\\
89	0\\
96	0\\
106	0\\
112	0\\
131	0\\
136	0\\
152	0\\
158	0\\
166	0\\
171	0\\
};
\addlegendentry{$-29^\circ$}

\addplot [color=green, mark=asterisk, mark options={solid, green}]
  table[row sep=crcr]{%
-170	0\\
-165	0\\
-157	0\\
-152	0\\
-143	0\\
-130	0\\
-110	0\\
-103	0\\
-90	0\\
-86	0\\
-83	0\\
-80	0\\
-74	0\\
-69	0\\
-61	0.682671143443601\\
-56	0\\
-47	0\\
-40	0\\
-34	1.59976485439274\\
-26	0\\
-21	0\\
-14	0.809875674116546\\
-7	2.28071032046856\\
0	2.83519813319267\\
6	2.38027868983561\\
10	1.44990817501026\\
13	2.3876826646563\\
16	0.996581610523315\\
22	2.18020381205686\\
27	0.762365748229149\\
35	0.702092982937745\\
40	0\\
49	1.14319677133967\\
56	0\\
62	0\\
70	0.597452256684282\\
75	0\\
82	0\\
89	0\\
96	0.997576949689843\\
102	0\\
106	0.691522844102605\\
109	0\\
112	0.937548154911093\\
118	0\\
123	0\\
131	0\\
136	0\\
145	0\\
152	0\\
158	0\\
166	0\\
171	0\\
178	0\\
};
\addlegendentry{$3^\circ$}
%

\end{axis}
\end{tikzpicture}%
	} \\
	\caption{Estimated data rate $T$ versus RX azimuth angles $\phi_{RX}$ for selected measurements and selected fixed BS orientations $\phi_{TX}$, representative of large-scale BS beam misalignment ($\theta_{RX}=0^\circ$ assumed throughout).\label{fig:fixedtx_linkop}}
\end{figure}

In real outdoor mm-wave network deployments, a misalignment at BS and UE is likely to occur simultaneously (\emph{cf.}~Fig.~\ref{fig:beammisscenarios}). We now study this case for the phased antenna array results, i.e. the beamsteering opportunities at the UE for a given BS beam misalignment. To this end, in Figs.~\ref{fig:rxssmisalignment}-\ref{fig:fixedtx_linkop} we show the estimated data rate $T$ over UE (RX) azimuth angles $\phi_{RX}$ for selected measurement scenarios and for selected representative BS (TX) orientations $\phi_{TX}$. In Figs.~\ref{fig:rxssmisalignment}-\ref{fig:fixedtx_linkop} we only consider the UE elevation angle $\theta_{RX}=0^\circ$, which yielded the highest data rates over all elevation angles (\emph{cf.}~Fig.~\ref{fig:rxelev}).

In Fig.~\ref{fig:rxssmisalignment} the BS orientations $\phi_{TX}$ were selected to study the effects of \emph{small-scale} misalignment in the order of the HPBW of around $6^\circ$. From Fig.~\ref{fig:rxssmisalignment} it is evident that the UE achieves the highest estimated data rate and has the maximum number of viable orientations when the transmitter is well aligned, i.e. $\phi_{TX}=0^\circ$. A small-scale misalignment of the transmitter of $\pm7^\circ$ produces a significant decrease in the maximum estimated data rate of up to 54\%. For a misalignment in the order of $\pm7^\circ$ at both BS and UE, the loss in estimated data rate is 38\%--100\% with a median of 70\% across all our results. Overall, we conclude that the joint effect of small-scale beam misalignment at BS and UE results in severe losses in data rate, placing the burden on agile beamtracking algorithms to update BS and UE orientation at a sufficient rate to ensure robustness of high data rate links for mobile users.




In Fig.~\ref{fig:fixedtx_linkop} the BS orientations $\phi_{TX}$ were selected to study link opportunities that occur due to sidelobes or independent secondary propagation paths, i.e. at orientations that deviate at \emph{large-scale} from the best orientation. In Fig.~\ref{subfig:fixedtx_linkop_1a}, for $\phi_{TX}=30^\circ$ the estimated data rates are below those for $\phi_{TX}=0^\circ$, but the shape of the curve is similar. This suggests that the TX has a sidelobe towards the RX when oriented at $\phi_{TX}=30^\circ$, using the same propagation path as for $\phi_{TX}=0^\circ$. By contrast, we observe an independent (NLOS) propagation path for $\phi_{RX}=-84^\circ$ and $\phi_{TX}=41^\circ$ which was also traced on the map in Fig.~\ref{subfig:paths_tx1}. Fig.~\ref{subfig:fixedtx_linkop_1c} reveals the same trend. The \mbox{$\phi_{RX}$-dependent} link opportunities for $\phi_{TX}=-37^\circ$ are similar to those for $\phi_{TX}=3^\circ$, suggesting that they resulted from a TX sidelobe. For $\phi_{TX}=-29^\circ$, a distinct signal path with a RX orientation of $\phi_{RX}=35^\circ$ yields the highest estimated data rate (\emph{cf.}~Fig.~\ref{subfig:paths_tx1} for the map). Based on Fig.~\ref{fig:fixedtx_linkop}, we conclude that identifying an independent secondary propagation path solely based on measured link quality over either AoD or AoA data is difficult, as sidelobes may also allow good data rates using the primary signal path. An approach of combining information from BS, UE, and external sources such as environmental awareness, e.g. \cite{simic_radmac:_2016}, may therefore be needed to identify independent secondary propagation paths that can be used to overcome blockage on the primary path.

\section{Discussion}
\label{sec:discussion}

Our mm-wave measurements in a German town center showed a limited number of 2--4 multipath clusters per receiver position, which is similar to the results reported for dense urban areas in modern metropolises \cite{rappaport_wideband_2015, samimi_28_2013, ko_millimeter-wave_2017}. This stands in contrast to expectations that the differences in building types and layout between a European town and modern metropolises would have an effect on the number of mm-wave multipath clusters. Our results thus suggest that mm-wave phased antenna array deployments are an attractive approach to achieve Gbps coverage in town centers at limited infrastructure cost thanks to their small area.

However, looking more closely at the structure of the received multipath clusters, we observe the paradigm shift that occurs when moving from horn antennas to real phased antenna arrays. With horn antennas, the clusters exhibit a smooth structure, giving a good indication of the actual underlying physical propagation paths. By contrast, with the phased antenna array the same physical propagation paths appear as irregular clusters due to the non-ideal radiation pattern with significant sidelobes and nulls, as shown in Figs.~\ref{subfig:16x2patternH}--\ref{subfig:16x2patternEl}. We also note that such strong sidelobes in the radiation pattern may result in significant intra-cell interference, which will have to be carefully considered during mm-wave network system design. 

As observed throughout this paper, the antenna radiation pattern plays a similarly crucial role in the effects of mm-wave beam misalignment. Our measurements with horn antennas show a monotonic relationship between misalignment and loss in received power, as also presented in earlier studies \cite{simic_60_2016, lee_field-measurement-based_2018}. By contrast, this does not hold for our phased antenna array results. For azimuth beam misalignments larger than the HPBW the degradation of estimated data rate is not monotonic with increasing beam misalignment for the non-ideal, phased antenna array radiation pattern. Instead, it varies irregularly with increasing beam misalignment. Consequently, measurement results obtained with horn antennas cannot simply be generalized to also hold for phased antenna arrays when investigating the effect of beam misalignment on the expected performance of real mm-wave networks. Similar implications hold for the theoretical works that make use of simplistic idealized antenna patterns, e.g. \cite{wildman_joint_2014}. 

Our results also showed that a small-scale beam misalignment at one station, i.e. either at BS or UE, caused a significant loss in estimated data rate that cannot be mitigated fully by the other station. We expect such a one-sided misalignment to be particularly harmful in uplink situations. As mm-wave UEs will likely not be able to feature phased antenna arrays of the same size as BSs, their antenna gain will be more limited due to the smaller number of antenna elements. Moreover, due to the limited battery capacity of the UE, also its output power will be more limited than the BS's output power. While the link performance could still be acceptable under beam misalignment when the BS transmits with high EIRP, this may not be the case under beam misalignments in uplink scenarios, where the UE EIRP is likely to be significantly lower. Thus, we expect to see a large asymmetry between uplink and downlink performance in mm-wave networks. Overall, we conclude that both the BS and UE in mm-wave networks will need to frequently adjust their orientation to counteract even one-sided small-scale beam misalignments. 

Furthermore, we note that one of the key challenges for enabling mm-wave outdoor networks is the handling of dynamic blockages by using an independent secondary unblocked signal path, as e.g. pedestrian blockage typically only concerns one multipath cluster and other multipath clusters remain unaffected \cite{weiler_measuring_2014}. However, due to the strength of sidelobes in non-ideal, phased antenna array radiation patterns, our results suggest that it may be difficult to distinguish between a true independent secondary propagation path and a link opportunity due to a sidelobe. This distinction is crucial, as link opportunities due to sidelobes on the primary propagation path are also affected by blockage of the primary propagation path. Beamsteering algorithms that obtain a list of viable link opportunities based on a simple RSSI search \cite{giordani_comparative_2016} are thus expected to suffer from this ambiguity if they do not restart at least a partial beam search upon link blockage.

Finally, our analysis suggest that, despite the wide elevation HPBW of our phased antenna array of $36^\circ$, 2D beamsteering is not sufficient to fully exploit available link opportunities at certain receiver positions. These results agree with the findings in \cite{zhou_following_2018} where it was demonstrated that 3D rotations can result in strong link degradations. This is particularly relevant as many beamsteering algorithms proposed in the literature only consider the planar case \cite{zhou_following_2018}. We expect that 3D beamsteering and hand-grip aware beamcombining \cite{alammouri_hand_2019} will be required to maintain a high data rate under rotation of the user device and different user activities. Despite the already increasing complexity of 5G-and-beyond protocol design, even additional external information, location-based environmental awareness, shared beamsteering information from devices in direct vicinity of a UE, or local data from a motion tracking sensor may further enhance the robustness and seamless coverage of mm-wave deployments under user mobility.

\section{Conclusions}
\label{sec:conclusion}

We presented the results of the first comprehensive large-scale outdoor mm-wave measurement study using a state-of-the-art phased antenna array in a European town. The data obtained over more than 5,000 systematic fine-grained AoA/AoD combinations per TX/RX pair was analyzed with respect to the number of available link opportunities and the effect of beam misalignment at the receiver and transmitter on the estimated mm-wave link data rate, and compared to reference measurements taken with a horn antenna. Our results show a limited number of 2--4 available multipath clusters per receiver location, indicating that the mm-wave multipath richness in a European town is surprisingly similar to that in dense urban metropolises as presented in prior literature. The results for the phased antenna array reveal that losses in the estimated data rate of up to 70\% occurred for small beam misalignments in the order of the HPBW, with significant and irregular variations in the estimated data rate for larger beam misalignments, caused by the non-ideal, phased antenna array radiation pattern. This stands in stark contrast to the horn antenna reference measurements, where the loss in estimated data rate was observed to be monotonically increasing with the misalignment error. Moreover, our results suggest that the characteristics of non-ideal, phased antenna array radiation patterns should be explicitly considered during the design and testing of mm-wave beamsteering algorithms. To this end, our ongoing work is focused on conducting further measurements in more diverse urban scenarios, explicitly studying the performance of beamsteering and beamtracking algorithms based on data obtained with real mm-wave phased antenna arrays.


\ifCLASSOPTIONcaptionsoff
  \newpage
\fi



\bibliographystyle{IEEEtran}
\bibliography{IEEEabrv,ms}

\end{document}